\documentclass[11pt,twoside,a4paper]{article}
\usepackage{hhline}
\usepackage{authblk}

\usepackage[utf8]{inputenc}
\usepackage[english]{babel}
\usepackage[colorlinks,linkcolor=blue,linktoc=page, citecolor=green, urlcolor=blue]{hyperref}
\usepackage{makeidx}
\usepackage[T1]{fontenc}
\usepackage{mathrsfs}
\usepackage{dsfont}

\usepackage{lmodern}
\usepackage[margin=0.8in]{geometry}
\usepackage[titletoc]{appendix}
\usepackage{amsfonts}
\usepackage{graphicx}
\usepackage{tcolorbox}
\tcbuselibrary{theorems}
\usepackage{lipsum}
\usepackage{marvosym} 
\usepackage{appendix}
\usepackage{ntheorem}
\usepackage{pbsi}
\usepackage[printonlyused]{acronym}
\usepackage{cancel}
\usepackage{empheq}

\usepackage{tikz}
\usetikzlibrary{positioning,arrows}
\usetikzlibrary{decorations.pathmorphing}
\usetikzlibrary{decorations.markings}

\usepackage{tabulary}
\usepackage{slashed}
\usepackage{esvect}
\usepackage{amsmath}
\usepackage{mathtools}

\usepackage{nccmath}
\usepackage{pifont}
\usepackage{colortbl}
\usepackage{xcolor}
\usepackage{cancel}
\usepackage{simplewick}
\usepackage{wrapfig}
\usepackage[font=small,labelfont=sc]{caption}
\usepackage{subcaption}

\renewcommand{\v}[1]{\ensuremath{\mathbf{#1}}} 
\newcommand{\gv}[1]{\ensuremath{\mbox{\boldmath$ #1 $}}} 
 
\newcommand{\grad}[1]{\gv{\nabla} #1} 
\renewcommand{\div}[1]{\gv{\nabla} \cdot #1} 
\let\baraccent=\= 
\renewcommand{\=}[1]{\stackrel{#1}{=}} 

\def\VecA{\mathbf{A}}
\def\VecV{\mathbf{v}}

\def\VecB{\mathbf{B}}

\def\VecJ{\mathbf{J}}

\def\TensD{\underline{\pmb{\text{D}}}}
\def\TensV{\underline{\pmb{\tau}}}
\def\TensK{\underline{\pmb{\kappa}}}

\newcommand{\pderiv}[2]{\frac{\partial #1}{\partial #2}}       

\begin{document}

\title{Simulations of COMPASS Vertical Displacement Events with a self-consistent model for halo currents including neutrals and sheath boundary conditions}

\renewcommand\Affilfont{\itshape\small}
\author[1]{F.J. Artola}
\author[1]{A. Loarte}
\author[5,6]{E. Matveeva}
\author[6]{J. Havlicek}
\author[6]{T. Markovic}
\author[6]{J. Adamek}
\author[6]{J. Cavalier}
\author[6]{L. Kripner}
\author[3,4]{G.T.A. Huijsmans}
\author[1]{M. Lehnen}
\author[2]{M. Hoelzl}
\author[6]{R. Panek}
\author[6]{the COMPASS team}
\author[7]{the JOREK team}
\affil[1]{ITER Organization, Route de Vinon sur Verdon, 13067 St Paul Lez Durance Cedex, France}
\affil[2]{Max Planck Institute for Plasmaphysics, Boltzmannstr. 2, 85748 Garching, Germany}
\affil[3]{CEA, IRFM, F-13108 St. Paul-lez-Durance cedex, France}
\affil[4]{Eindhoven University of Technology, 5612 AP Eindhoven, The Netherlands}
\affil[5]{Charles University, Faculty of Mathematics and Physics, Prague, Czech Republic}
\affil[6]{Institute of Plasma Physics of CAS, Za Slovankou 3, 182 00 Prague, Czech Republic}
\affil[7]{please refer to [M Hoelzl, G T A Huijsmans, S J P Pamela, M Becoulet, E Nardon, F J Artola, B Nkonga et al, Nuclear Fusion, in preparation] }
\date{\vspace{-3ex}}

\interfootnotelinepenalty=10000

\maketitle
Corresponding author: \textbf{A. Loarte}
\hspace{1.5cm} Email: \textcolor{blue}{alberto.loarte@iter.org}\\

\abstract{The understanding of the halo current properties during disruptions is key to design and operate large scale tokamaks in view of the large thermal and electromagnetic loads that they entail. For the first time, we present a fully self-consistent model for halo current simulations including neutral particles and sheath boundary conditions. The model is used to simulate Vertical Displacement Events (VDEs) occurring in the COMPASS tokamak.     Recent COMPASS experiments have shown that the parallel halo current density at the plasma-wall interface is limited by the ion saturation current during VDE-induced disruptions. We show that usual MHD boundary conditions can lead to the violation of this physical limit and we implement this current density limitation through a boundary condition for the electrostatic potential. Sheath boundary conditions for the density, the heat flux, the parallel velocity and a realistic parameter choice (e.g. Spitzer $\eta$ and Spitzer-Härm $\chi_\parallel$ values) extend present VDE simulations beyond the state of the art. Experimental measurements of the  current density, temperature and heat flux profiles at the COMPASS divertor are compared with the results obtained from axisymmetric simulations. Since the ion saturation current density ($J_{sat}$) is shown to be essential to determine the halo current profile, parametric scans are performed to study its dependence on different quantities such as the plasma resistivity and the particle and heat diffusion coefficients. In this respect, the plasma resistivity in the halo region broadens significantly  the $J_{sat}$ profile, increasing the halo width at a similar total halo current.

\medskip

\section{Introduction}
The operation of large scale tokamaks such as ITER and DEMO requires a Disruption Mitigation System (DMS) to minimize the impact of disruption loads on the plasma facing components and on the vacuum vessel \cite{lehnen2015disruptions}. The design and optimization of such a system must be guided by experiments and validated numerical codes. In this respect, high fidelity simulations should be able to predict the worst case scenarios in terms of heat and electromagnetic loads in order to quantify the load reduction provided by the different mitigation schemes. 

\medskip

During disruptions, currents flowing along the open field lines (halo currents) can produce large forces on the vacuum vessel and also large heat loads on the plasma facing components by converting the plasma magnetic energy into thermal energy via Ohmic heating. The parametric dependencies of the current density and the heat flux profiles during disruptions are not yet well understood. MHD simulations for disruptions are numerically challenging and therefore previous simulations have typically oversimplified the boundary conditions for different plasma quantities (e.g. Dirichlet conditions for temperature and density). 

\medskip

In this paper we present, for the first time,  a fully self-consistent model for halo current simulations including neutral particles and a set of advanced boundary conditions.  Sheath boundary conditions for the density, the heat flux and the parallel velocity together with a realistic parameter choice (e.g. Spitzer $\eta$ and Spitzer-Härm $\chi_\parallel$ values) extend non-linear MHD VDE simulations beyond the state of the art. Additionally we introduce a boundary condition for the electrostatic potential in order to limit the halo current density to the ion saturation current. This limitation comes from basic sheath plasma physics and has recently been observed in COMPASS  disruption experiments \cite{adamek2020}. Neutral particles are required to calculate the evolution of the boundary density, which is key to determine the ion saturation current density. As a first step, the neutral particle density is modeled with a fluid continuity equation and an effective diffusion coefficient. For this work we use the JOREK code to perform axisymetric MHD simulations of COMPASS VDEs and compare the modeled results with experimental measurements. Since the ion saturation current density ($J_{sat}$) will be shown to be essential to determine the halo current profile, parametric scans are performed to study its dependence on different quantities such as the plasma resistivity and the particle and heat diffusion coefficients. Recent simulations with the NIMROD code have explored the effect of different boundary conditions for the plasma density, temperature and velocity on VDEs \cite{bunkers2020}. However the important effects of neutral particles and the ion saturation current limitation were not included.

\medskip

This paper is organized as follows: In section \ref{sec:model} the MHD model and the boundary conditions  are explained in detail. The numerical setup is presented in section \ref{sec:setup} together with the parameter choice used to simulate the COMPASS VDE experiments. The large effect of the sheath boundary conditions on the evolution of the halo current is discussed in \ref{sec:influence_BC}. In that section, the influence of plasma viscosity, neutral particle reflection and the implemented minimum heat and particle fluxes are also discussed. The comparison of the simulations with the experimental results is given in section \ref{sec:comparison}. The dependencies of the halo current profile on different parameters is researched in section \ref{sec:sensitivity}. Finally the conclusions are summarized in section \ref{sec:conclusions}.

\section{Model and boundary conditions}
\label{sec:model}
The implicit non-linear  code JOREK solves the extended magneto-hydrodynamic (MHD) equations in realistic tokamak geometries including open field line regions~\cite{huysmans2007mhd,overview}. In JOREK  the poloidal plane is discretized with quadrilateral bicubic Bezier elements using an isoparametric mapping~\cite{Czarny2008Jocp}. Fourier harmonics are used to decompose the toroidal direction and for the time stepping  typically  the Crank-Nicholson  or the BDF2 Gears scheme are used. The code contains a hierarchy of different MHD models with various extensions. The MHD model used for this work includes an equation for fluid neutral particles   
\begin{align}
\pderiv{\VecA}{t} &= \VecV\times\VecB - \eta\VecJ  - \nabla \Phi, \label{eq:mhd:A}\\
\rho\pderiv{\VecV}{t} &= -\rho\VecV\cdot\nabla\VecV - \nabla p +
\VecJ\times\VecB + \nabla\cdot\TensV 
- (\rho\rho_n S_{ion}-\rho^2 \alpha_{rec})\VecV,
\label{eq:mhd:v} \\
\pderiv{\rho}{t} &=
-\nabla\cdot(\rho\,\VecV)
+\nabla\cdot(\TensD\nabla\rho)
+(\rho\rho_n S_{ion}-\rho^2 \alpha_{rec})
,\label{eq:mhd:rho}
\\
\pderiv{\rho_n}{t} &=
\nabla\cdot(\TensD_n\nabla\rho_n)
-(\rho\rho_n S_{ion}-\rho^2 \alpha_{rec})
,\label{eq:mhd:rho_n}
\\
\pderiv{p}{t} &= -\VecV\cdot\nabla p
- \gamma p\nabla\cdot\VecV
+ \nabla\cdot(\TensK\nabla T) + (\gamma -1)\eta J^2  \nonumber  \\
  & \hspace*{2cm}\rule{0mm}{4mm} 
  -\xi_{ion}\rho\rho_n S_{ion} - \rho\rho_n L_{lines} - \rho^2 L_{brem}
\label{eq:mhd:p}
\end{align}
with the following reduced MHD ansatz for the magnetic field ($\VecB$) and the mean plasma velocity ($\VecV$)
\begin{align}%
\VecB &= \nabla\psi\times\nabla\phi + F_0 \nabla\phi, \label{eq:B}\\
\VecV &= -\frac{R^2}{F_0}\nabla\Phi\times\nabla\phi +\VecV_\parallel, \label{eq:v}
\end{align}
where $\psi$ is the poloidal magnetic flux and $F_0 = RB_\phi$ is a constant representing the main reduced MHD assumption, which is that the toroidal field is fixed in time. Note however that poloidal currents are not fixed in time and evolve according to the current conservation and momentum balance equations, but their contribution to the toroidal field is neglected due to the large vacuum field. A recent benchmark with full MHD codes has demonstrated that halo currents can be described accurately in this way \cite{krebs2020axisymmetric}. The quantities shown in equations (1)-(7) are  the magnetic potential ($\VecA$), the ion density ($\rho$), the neutral density ($\rho_n$), the total pressure ($p$), the total temperature ($T\equiv T_e + T_i$), the electrostatic potential ($\Phi$) and the current density ($\VecJ$). The parameters related to the neutral particle fluid are the ionization and recombination rates ($S_{ion}$, $\alpha_{rec}$), the Deuterium ionization energy ($\xi_{ion}$) and the line ($L_{lines}$) and bremsstrahlung ($L_{brem}$) radiation coefficients. However the charge exchange process is not included, which leads to momentum losses at low temperatures ($T\sim 1-10$ eV) and will be included in future works. Other parameters in equations (1)-(7) are the plasma resistivity ($\eta$), the stress tensor($\TensV$), the thermal  conductivity and the particle diffusion coefficients ($\TensK,\TensD,\TensD_n$) and the ratio of specific heats ($\gamma$). The thermal conductivity coefficients $\TensK$ present a high anisotropy (i.e. $\kappa_\parallel \gg \kappa_\perp$) while the particle diffusion coefficients are normally isotropic. Note that the reduction of the equations is ansatz-based, conserves energy \cite{franck2014energy} and does not result from geometrical ordering assumptions. The effect of impurity radiation plays an important role in the disruption dynamics, however this effect is not  included in the present simulations and it is left for future work.

\medskip

The resistive wall and the free-boundary conditions are included by coupling JOREK to the STARWALL code \cite{merkel2015linear,holzl2012coupling, artolasuch:tel-02012234}. The coupling is performed by solving the full vacuum domain with a Green's function method and therefore the vacuum does not need to be meshed.  STARWALL uses the thin wall assumption  and discretises the wall with linear triangular elements. Although the employed wall formalism is  implemented for 3D thin walls including holes, the  setup used here is restricted to an axisymmetric wall.

\subsection*{Boundary conditions}
In this subsection, the boundary conditions used throughout this paper are explained. The free-boundary conditions for the poloidal magnetic flux are explained in detail in the references \cite{holzl2012coupling, artolasuch:tel-02012234}.

\subsubsection*{Parallel velocity ($v_\parallel$)}
The employed boundary condition for the parallel velocity to the magnetic field is a special case of the Chodura-Riemann condition \cite{stangeby1995} at the magnetic pre-sheath
\begin{equation}
	v_\parallel = g(b_n) c_s
\end{equation}
where $c_s \equiv \sqrt{\gamma k_B T / m_i}$ is the ion sound-speed, and $g(b_n)$ is a function of the normal projection of the magnetic field into the wall ($b_n\equiv \v{B}\cdot\v{n}/B$). Here $\v{n}$ is a unit vector which is perpendicular to the boundary and points outside the plasma domain. The function $g$ is needed to ensure that ions always flow towards the wall ($\v{v}_\parallel\cdot\v{n}> 0$) and that the variable $v_\parallel$ has a smooth transition in the regions where $b_n$ changes sign. Otherwise the condition ($\v{v}_\parallel\cdot\v{n}> 0$) leads to discontinuities in $v_\parallel$ along the boundary that cannot be resolved (e.g. $v_\parallel$ would jump from $c_s$ to $-c_s$). The chosen form of the $g$ function is
\begin{equation}
g(b_n) = \textrm{sign}(b_n) \left[ \frac{1}{4}(1+\tanh ((|b_n| - g_1)/g_2)^2 - g_3\right]
\end{equation}
and for this work the chosen coefficients are $(g_1,g_2,g_3)=(0.02,0.016,0.005754)$ such that $g(0)=0$ and $g\approx 1$ when the angle of incidence is larger than $2^{\circ}$.

\subsubsection*{Heat flux}
The boundary condition for the heat flux is based on reference \cite{stangeby2000plasma}
\begin{equation}
\v{q}\cdot\v{n}=\gamma_{sh} \, n \, k_B T_e\,  \v{v}_\parallel\cdot\v{n} + q_{min}
\label{eq:q_flux_BC}
\end{equation}
where $n$ is the particle density and $\gamma_{sh}$ is the sheath transmission coefficient. In this work the value $\gamma_{sh}=11$ is chosen as it has been found in COMPASS through experimental observations for steady state plasmas \cite{Horacek_2020}. Note that the evolution of $\gamma_{sh}$ during disruptions is presently unknown. Since the boundary conditions for the parallel flow are such that $v_\parallel = 0$ when the field lines are tangential to the wall (e.g. limiter point), a minimum heat flux ($q_{min}$) has been introduced. Otherwise, if $\v{q}\cdot\v{n}=0$ at the tangency points, the thermal energy would accumulate artificially at the boundary. We choose the minimum heat flux as $q_{min}=\gamma_{sh}n_e k_B T_e c_s \sin (\vartheta_{min})$, where unless explicitly stated, $\vartheta_{min}=2^\circ$. The choice of a minimum value is also motivated by the reference \cite{matthews1990investigation}, in which significant particle and heat fluxes are found due to significant radial diffusion of energy and particles from the plasma at grazing angles. 

\subsubsection*{Ion flux}
The employed boundary condition for the ion flux implies that ions leave the plasma domain at the parallel velocity
 
\begin{equation}
\v{\Gamma}\cdot\v{n}= n \v{v}_\parallel\cdot\v{n} + \Gamma_{min}
\label{eq:part_flux_BC}
\end{equation}
Similar to the heat flux boundary condition, we introduce a minimum particle flux to avoid artificial accumulation of particles, $\Gamma_{min}= n c_s \sin (\vartheta_{min})$. 

\subsubsection*{Neutral particle flux}
In order to ensure conservation of particles through the simulation the neutral particle flux is
\begin{equation}
\v{\Gamma}_n\cdot\v{n} = - \v{\Gamma}\cdot\v{n}
\end{equation}
which implies that all ions leaving the boundary come back as neutral particles.

\subsubsection*{Electrostatic potential}
Basic physics of the plasma implies that, in the presence of a sheath between the plasma and the material surfaces in contact with it, the maximum plasma current density that can flow along the field line is limited by the ion saturation current. Therefore in the presence of sufficiently large voltages induced by the disruption dynamics, the halo current could be directly determined by the ion saturation current. This hypothesis has been confirmed in recent COMPASS VDE experiments \cite{adamek2020}. This was achieved by measuring independently the current flowing into the divertor through grounded Langmuir probes (to the vacuum vessel) and the ion saturation current at nearby locations with biased Langmuir probes. During disruptions, the parallel current density ($\v{J}_\parallel$) was similar to the measured ion saturation current ($\v{J}_{sat}$). As we will show later, MHD models do not normally include this limitation of the current density and can violate it. For that reason we explore a suitable boundary condition for the MHD code JOREK including that limitation. The parallel current integrated over a field line  has the following evolution in reduced MHD
\begin{equation}
\int_a^b \eta J_\parallel dl = -(\Phi_b - \Phi_a) - \int_a^b \frac{B_\phi}{R|B|}\pderiv{\psi}{t} dl  
\end{equation}
where $a$ and $b$ are respectively the starting and ending points of the field line and $dl$ is the length differential along the field line. The latter relation is found by integrating the parallel electric field along the field line. For a given flux variation (e.g. caused by the decay  of the plasma current)  the current density along the field line is limited by the potential at its end points $(\Phi_b, \Phi_a)$. Therefore the current density can be controlled by applying boundary conditions for $\Phi$. At the plasma sheath, analytical expressions relating $\Phi$ and $J_\parallel$ that feature such a limit are well known from Langmuir probe theory \cite{stangeby2000plasma}. The implementation of a boundary condition based on the Langmuir formulation is numerically very challenging and thus we have used a simplified  boundary condition to limit the current to the ion saturation current

\begin{equation}
\Phi =
\begin{cases} 
      \Phi_w + \alpha(\v{J}_\parallel-\v{J}_{sat})\cdot\v{n},  &\textrm{if} \quad \v{J}_\parallel\cdot\v{n} \ge \v{J}_{sat}\cdot\v{n} \\
      \Phi_w, &\textrm{if} \quad \v{J}_\parallel\cdot\v{n} < \v{J}_{sat}\cdot\v{n}
\end{cases}  
\end{equation}
where $\alpha$ is a constant. The boundary condition is such that when the parallel current ($\v{J}_\parallel$) exceeds the ion saturation current ($\v{J}_{sat}$), a voltage is driven with respect to the wall in order to reduce $\v{J}_\parallel$. By choosing a large $\alpha$ it is ensured that the parallel current remains close to $\v{J}_{sat}$. If  $\v{J}_\parallel\cdot\v{n} < \v{J}_{sat}\cdot\v{n}$ the potential is set to the wall potential, which is chosen to be $\Phi_w=0$  (the boundary acts as a perfect conductor in the poloidal direction).

\medskip

The boundary condition for the potential can be expressed in terms of the JOREK variables. Assuming force-free balance in the open field- line region ($\v{J}\times\v{B}=0$), the parallel current density is $\v{J}_\parallel = -j \v{B} /F_0$ with $j\equiv -RJ_\phi$ being the JOREK variable. The ion saturation current is $\v{J}_{sat}=\text{sign}(b_n) \, e \,n \,c_s\, \v{b} $ and therefore the situation $\v{J}_\parallel\cdot\v{n} = \v{J}_{sat}\cdot\v{n}$ corresponds to $j=j_{sat}\equiv -\text{sign}(b_n) \,e\, n \,c_s\, F_0/ |B|$. Finally the JOREK boundary condition is 

\begin{equation}
\Phi =
\begin{cases} 
      -\frac{\alpha}{F_0}(j-j_{sat})\v{B}\cdot\v{n},  &\textrm{if} \quad \v{J}_\parallel\cdot\v{n} \ge \v{J}_{sat}\cdot\v{n} \\
      0, &\textrm{if} \quad \v{J}_\parallel\cdot\v{n} < \v{J}_{sat}\cdot\v{n}
\end{cases}
\label{eq:phi_jsat}
\end{equation}

\subsubsection*{Vorticity and current density}
In order to avoid second order derivatives in the code, auxiliary variables such as the toroidal vorticity $w_\phi = \frac{1}{F_0}\Delta \Phi$ and the toroidal current density $j= \Delta^* \psi$ are introduced. The boundary condition for the vorticity is simply $w_\phi=0$ and $j$ evolves according to Ampere's law ($j= \Delta^* \psi$).

\subsubsection*{Implementation}
The heat and particle fluxes are implemented as natural boundary conditions of the finite element method. When performing the weak form and the partial integration of the $\TensD$ and the $\TensK$ terms, the latter expressions for the fluxes are replaced in the arising boundary integrals. When Dirichlet boundary conditions are applied, a penalization method is used. The method can be interpreted as  adding a boundary integral term to the governing equations. For example the following term is added to the parallel flow equation
\begin{equation}
\oint Z v^*(v_\parallel - g(b_n) c_s) dS
\end{equation}
where $v^*$ is a test function of the FEM, $dS$ is the boundary surface and $Z=10^{12}$ is an arbitrarily large value used to enforce the boundary condition.

\begin{figure}
  \includegraphics[width=1\textwidth]{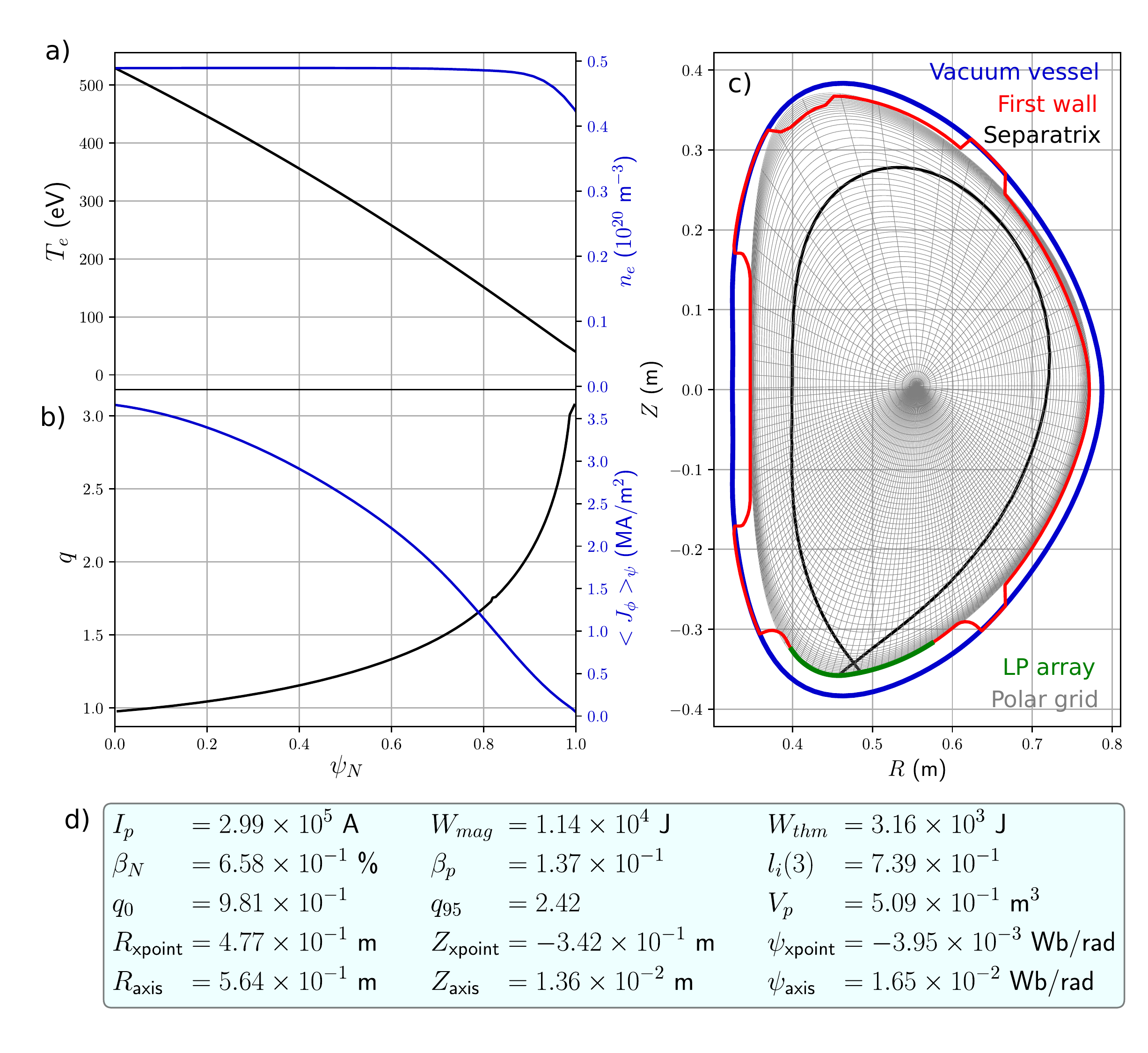}
\caption{Plasma configuration obtained after the steady state run in JOREK for the COMPASS shot \#19172 at $t=1090.0$ ms. (a) Temperature and density profiles at the midplane, (b) safety factor and averaged toroidal current density profiles as functions of the normalized poloidal flux. (c) Geometry of the COMPASS vacuum vessel and first wall, the Langmuir probe array location, the plasma separatrix and the polar grid used in JOREK. (d) List of  relevant scalars describing the equilibrium.}
\label{fig:eq_plot}
\end{figure}

\section{Simulation setup}
\label{sec:setup}

\subsection{Equilibrium and steady state}
In this subsection, we describe the plasma state prior to the VDE simulation. JOREK solves the Grad-Shafranov equation in order to obtain an initial condition that is accurate for the employed finite element method. In its free-boundary version, the computation of the equilibrium requires to specify the pressure and the toroidal current profiles as well as the geometry and the currents of the PF coils. An initial equilibrium was computed based on an EFIT reconstruction of the COMPASS discharge \#19172 at $t=1090.0$ ms. As a kinetic version of EFIT was not available, the pressure profile  was represented by a parabolic function. Similarly, the current profile is also parabolic, with the constraints that $q_\textrm{axis}\sim 1$ and that the final total current is the experimental one ($I_p=300$ kA). The currents flowing in the 5 independent circuits of the COMPASS PF coils are $(I_\textrm{BR}, I_\textrm{BV}, I_\textrm{EFPS}, I_\textrm{MFPS}, I_\textrm{SFPS}) = (0.20,2.00,-13.2, -7.16, -9.42)$ kA/turn. The $I_\textrm{BR}$ circuit is used for feedback control on the vertical  position and its current has been increased to 0.9 kA/turn in the simulation to obtain a more accurate matching of the vertical position of the magnetic axis ($Z_\textrm{axis}$).

\medskip

For the plasma computational domain a polar grid has been chosen (see figure \ref{fig:eq_plot} (c)). The boundary of the grid matches the COMPASS first wall in the regions of main plasma-wall interaction. Its simplicity facilitates the implementation of the boundary conditions but some regions are not represented accurately. The grid features radial mesh accumulation at the computational boundary in order to resolve large temperature gradients that may arise. The vacuum vessel (blue curve in figure \ref{fig:eq_plot} (c)) is an axisymmetric version of the COMPASS wall and it is discretized in STARWALL with triangular elements using the thin wall approximation. The toroidal resistance of the wall is 40\% smaller than the value given by COMPASS specifications ($\mathcal{R}_w = 0.63 \textrm{ m}\Omega$) in order to have a better matching of the vertical position evolution with the experiment (see next section).

\medskip

Once the initial condition is given, an axisymmetric time evolution simulation (of $\sim 4$ ms) is performed in order to obtain a steady state plasma. The parameters that have been used for this phase are summarized in table \ref{tab:param_steady}. Note that we use temperature dependent $\eta$ and $\kappa_\parallel$ with their corresponding Spitzer and Spitzer-Härm values. A current source  term ($\eta j_0$) is added to the magnetic flux equation in order to fix the initial current profile over time ($\partial_t\psi=...+ \eta (j - j_0)$. In the core, the steady state is reached when the Ohmic heating and the conductive perpendicular transport are balanced. The chosen $\kappa_\perp$ is such that the final thermal energy ($W_{th}=3.2$ kJ) is similar to the energy obtained when mapping the Thomson scattering ($n_e, T_e$) measurements  into the EFIT reconstruction assuming $T_i = 0.5 T_e$. The  neutral diffusion coefficient was taken from JOREK simulations without charge exchange that were compared to SOLPS-ITER simulations \cite{Smith_2020}. The ion and neutral particle densities are in equilibrium when the diffusive transport and the ionization and recombination processes are balanced. The final core profiles are shown in figure \ref{fig:eq_plot} (a) and (b). With the chosen parameters, the electron density accumulates at the High Field Side (HFS) near the lower X-point (see Figure \ref{fig:eq_neTemap} right), which explains the increasing plasma density towards the separatrix seen in Figure \ref{fig:eq_plot} (a). The final plasma temperature profile is also shown in Figure \ref{fig:eq_neTemap} (left) and several parameters describing the final steady state equilibrium are presented in Figure \ref{fig:eq_plot} (d).

\begin{table}[ht]

\small
\def\arraystretch{1.5}
\centering
\resizebox{0.85\textwidth}{!}{%
\begin{tabular}{|c|c|c|}
\hline

\textbf{Parameter} & \textbf{Value}   &  \textbf{Description}\\ \hhline{|=|=|=|}
$ D$ & $  4.57 \textrm{ m}^2/\textrm{s}$ & Isotropic particle diffusion coefficient \\ \hline
$D_n$ & $ 228 \textrm{ m}^2/\textrm{s}$ & Isotropic neutral particle diffusion coefficient \\ \hline
$\kappa_\perp$ & $  0.988\times 10^{20} \textrm{ (m s)}^{-1}$ & Perpendicular thermal conductivity \\ \hline
$\kappa_\parallel = \kappa_0 (T_e/T_{e0})^{5/2} $ & $ \kappa_0 = 3.73\times10^{29}  \textrm{ (m s)}^{-1}$ & Parallel thermal conductivity  \\ \hline
$\eta = \eta_0 (T_e/T_{e0})^{-3/2} $ & $ \eta_0 = 4.19\times10^{-8} \Omega\textrm{ m}$ & Resistivity \\ \hline
$(\mu_\parallel, \mu_\perp)$ & $ (3.48, 1.05) \times 10^{-7} \textrm{ kg}/\textrm{(m s)}$ & 
Parallel/perpendicular dynamic viscosity \\ \hline
$\gamma_{sh} $ & $ 11 $& Sheath transmission coefficient \\ \hline

\end{tabular}%
}

\caption{Parameters used during the steady state run. For the temperature dependent parameters the reference temperature is $T_{e0}= 1 $ keV. Note that except for $\eta$ and $\kappa_\parallel$ the coefficients are spatially constant. }
\label{tab:param_steady}
\end{table}

\begin{figure}
\centering
  \includegraphics[width=0.7\textwidth]{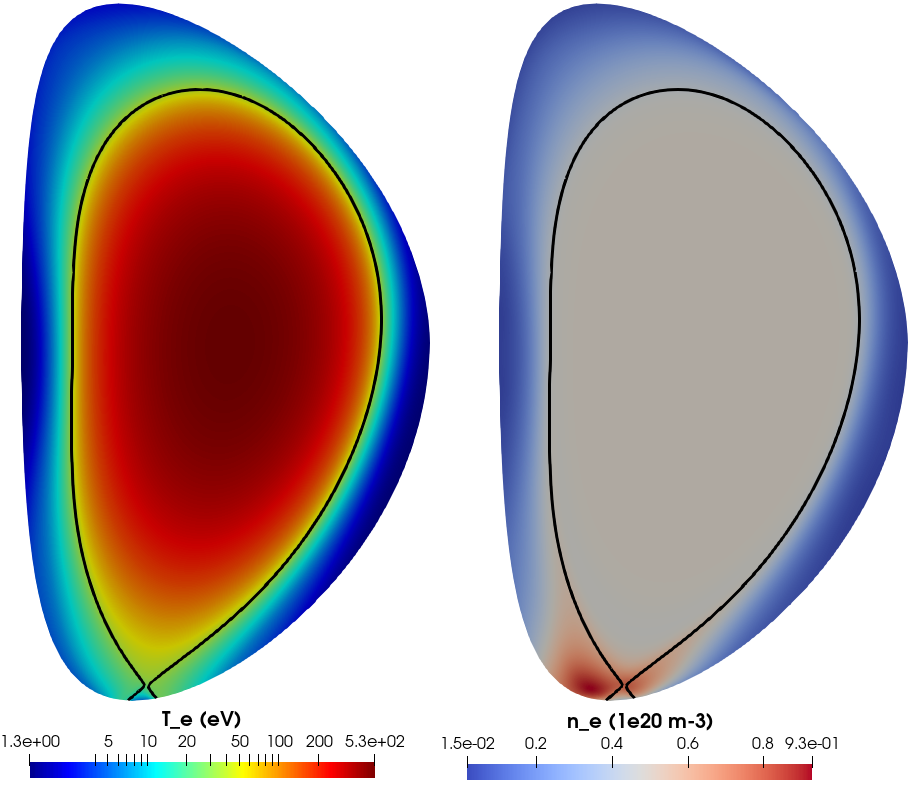}
\caption{Electron temperature distribution in logarithmic scale (left) and electron density distribution in linear scale (right) at the end of the steady state run. The separatrix is indicated by the black curve.}
\label{fig:eq_neTemap}
\end{figure}

\begin{figure}
\centering
  \includegraphics[width=0.7\textwidth]{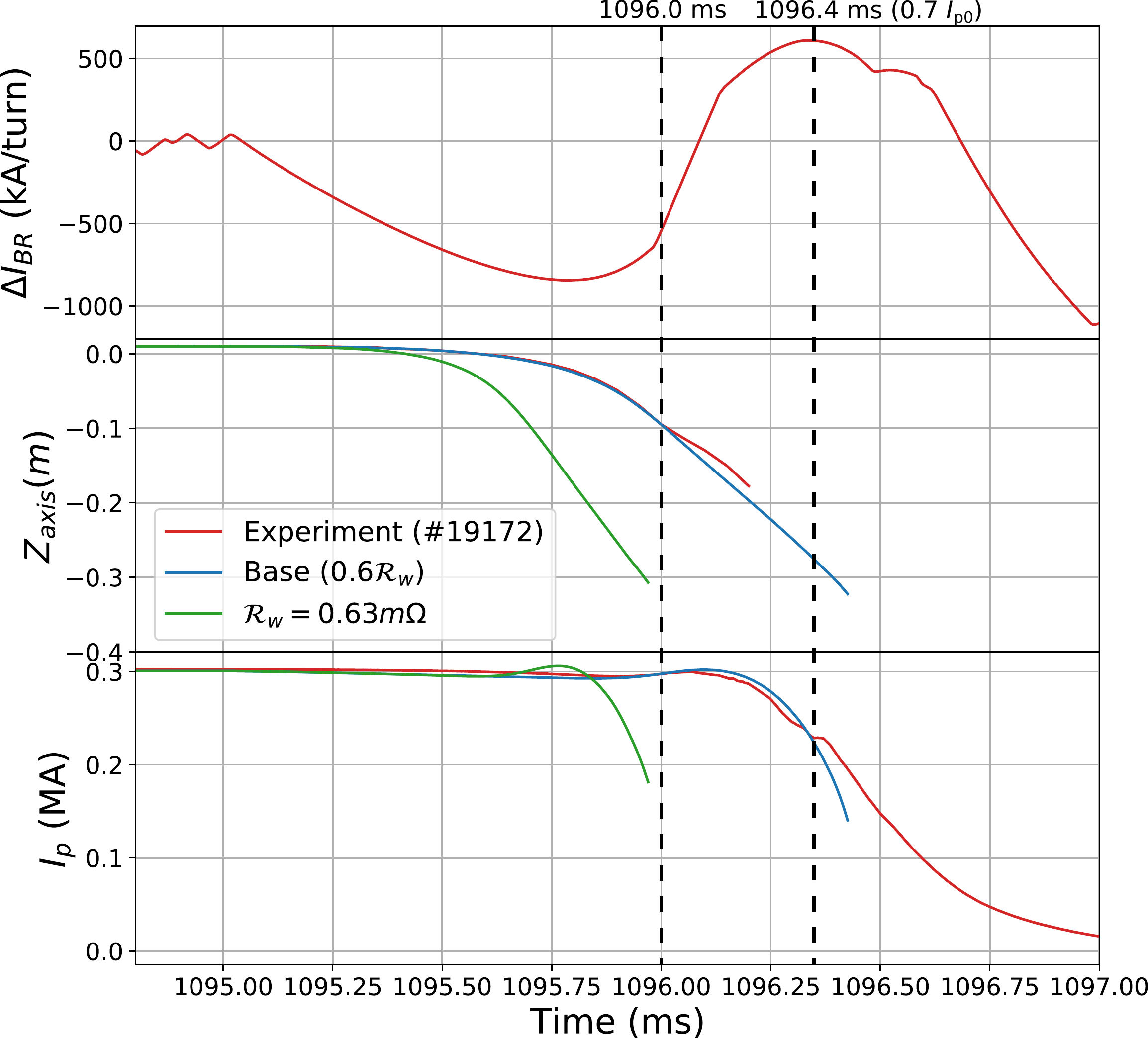}
\caption{Comparison of two JOREK runs with different wall resistances (blue and green lines) and the experimental time traces (red). (Top) Current change in the $I_\text{BR}$ circuit. (Middle) Vertical position of the magnetic axis. (Bottom) Total plasma current inside the plasma domain. The vertical dashed lines indicate the two times ($t=1096.0$ ms) and ($t=1096.4$ ms) that are chosen in sections \ref{sec:comparison} and \ref{sec:sensitivity} to plot the divertor profiles.}
\label{fig:kick_etaw}
\end{figure}

\begin{figure}
\centering
  \includegraphics[width=0.65\textwidth]{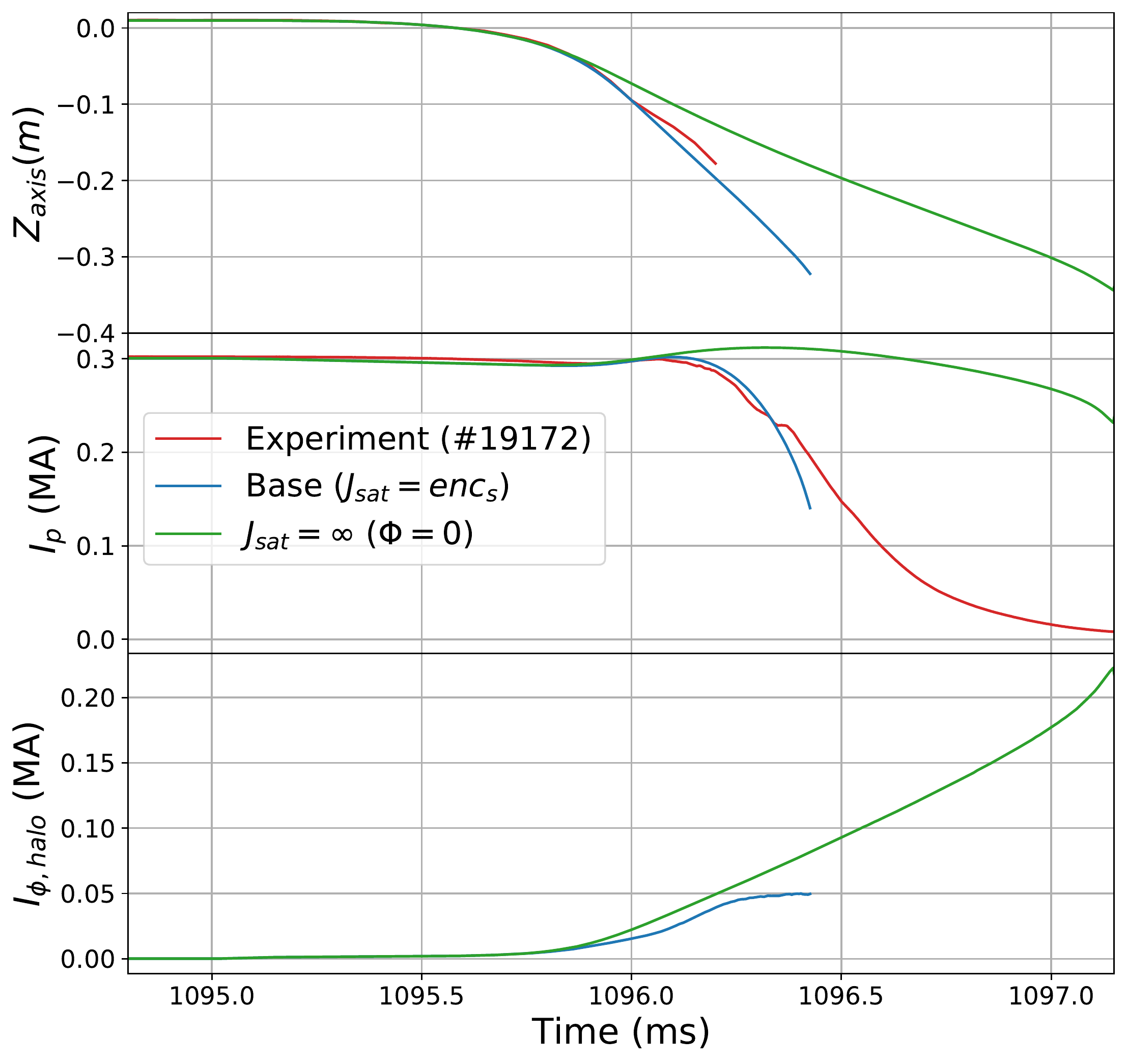}
\caption{ Comparison of two simulations with a different $J_{sat}$ value used for the boundary condition given by expression \ref{eq:phi_jsat}. (Top) Vertical position of the magnetic axis. (Middle) Total plasma current. (Bottom) Toroidal plasma current in the halo region. The figure shows the stabilizing effect of the halo currents and the large influence of the boundary condition for $\Phi$ for the dynamics of the VDE. }
\label{fig:compare_u0_dyn}
\end{figure}

\subsection{Triggering of the VDE}
During the steady state run, vertical position control is achieved with a PID controller acting on the $I_\text{BR}$ PF coil circuit. In the experiments a vertical kick is performed by applying a current wave-form to the $I_\text{BR}$ circuit in order to force a downwards VDE. In the simulation the same PF coil current variation ($\Delta I_\text{BR}$) as in the experiment is specified and leads to a similar vertical position evolution (see figure \ref{fig:kick_etaw}). We remind the reader that we have reduced the wall resistance by 40\% in order to have a better matching with the experiment. The case with the COMPASS real wall resistance is also present in figure \ref{fig:kick_etaw} showing  a faster plasma displacement modelled by JOREK than measured in the experiment. For the VDE phase, the employed parameters were not modified with respect to the steady state phase. The main difference is that the current source term is switched-off and the current is allowed to decay. Also the code is run in axisymmetric mode.

\section{Influence of boundary conditions}
\label{sec:influence_BC}
\subsection{Electric potential and ion saturation current}
\label{sec:influence_BC_Phi}
The boundary condition given by expression \ref{eq:phi_jsat} is tested in this subsection and compared to the usual boundary condition applied in MHD simulations. When the usual MHD boundary condition for the electrostatic potential $\Phi=0$ is applied, large halo currents are induced in the SOL and the decay time of the plasma current is much longer than observed in the experiment (see green and red lines of Figure \ref{fig:compare_u0_dyn}). In this case  the halo current has a very large width and presents a non-monotonic profile as indicated by Figure \ref{fig:compare_u0} b), in which, the toroidal current density is shown at a late time of the VDE. A broad halo width ($w_{halo}$) decreases the total resistance of the halo region ($\mathcal{R}_{halo}\sim \eta_{halo}/w_{halo}$) and consequently the total plasma current decays at a slower rate.  Since in our simulations the core and the halo regions are  treated with the same formalism, the full plasma domain acts as a conductor with resistivity $\eta(T_e)$. Due to the electric field created by the moving plasma core, currents appearing in the far SOL (Figure \ref{fig:compare_u0} b) can be self-sustained through Ohmic heating (which increases $T_e$ and makes the region more conductive). 

\begin{figure}
\centering
  \includegraphics[width=0.75\textwidth]{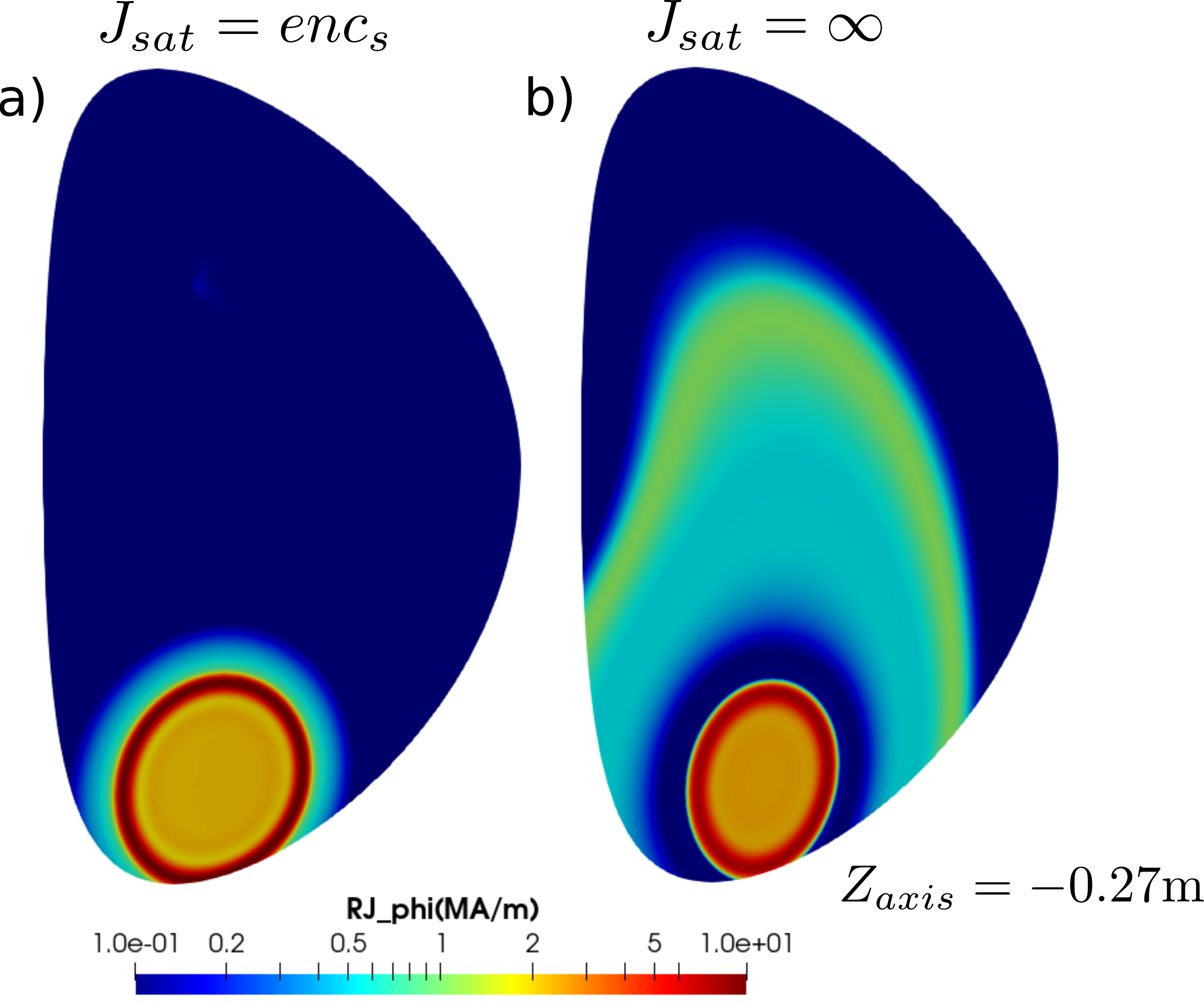}
\caption{Toroidal current density at $Z_{axis}=-0.27$ m for the same two simulations as in Figure \ref{fig:compare_u0_dyn} with different boundary conditions for the electrostatic potential. (a) Simulation with $J_{sat}=enc_s$  and $\Phi$ employed in expression \ref{eq:phi_jsat}. (b) Simulation with $\Phi=0$ as boundary condition (equivalent to use $J_{sat}=\infty$ in expression \ref{eq:phi_jsat}). The figure shows the large influence of the boundary condition for $\Phi$ on the distribution of the halo currents.  }
\label{fig:compare_u0}
\end{figure}

\medskip

The case with the boundary condition $\Phi=0$ leads to non-physical results since the normal current density ($\v{J}\cdot\v{n}$) largely exceeds the ion saturation current ($\v{J}_{sat}\cdot\v{n}$) at different locations along the plasma-wall interface (see Figure \ref{fig:Jlim_violation}). In order to visualize these quantities along the boundary of the plasma domain, the polar coordinate $\theta_w$ is introduced. The polar coordinate system is centered at $(R,Z)=(0.56,0)$ m such that $\theta_w=0$ corresponds to the outer midplane ($Z=0$). The results shown in Figure \ref{fig:Jlim_violation} imply that the usual boundary condition ($\Phi=0$) allows large currents to flow at locations with very low plasma densities (e.g. at $\theta_w/2\pi=0.6$ near the high field side midplane).

\medskip

When the boundary condition given by equation \ref{eq:phi_jsat} is applied for the electrostatic potential, the halo current is limited by the ion saturation current. Since the limitation is given by $J_{sat}\equiv enc_s$, the halo current is mainly constrained by the particle density profile ($n$) which evolves according to ionization, recombination, diffusion and convection. In practice, this implies that the halo current cannot be induced in the far SOL regions where the particle density is low. The latter can be observed in Figure \ref{fig:compare_u0} a) as well as in Figure \ref{fig:Jlim_proof}. As a consequence, the total induced halo current is much smaller than in the case without current limitation (see blue and  green lines of Figure \ref{fig:compare_u0_dyn}). The halo current width is also severely reduced (see Figure \ref{fig:compare_u0} a)) and thus the total plasma current decays at a faster rate, which is closer to experimental observations. Therefore the performed simulations show that the boundary condition for the electrostatic potential ($\Phi$) plays a very significant role for both plasma dynamics and distribution of the current profile in the halo region (see Figure \ref{fig:Jlim_proof}).

\medskip

The boundary condition shown in equation \ref{eq:phi_jsat} provides an upper limit only for the positive current ($\v{J}\cdot\v{n}>0$), which in this case is  induced by the VDE at the High Field Side (HFS). In the Low Field Side (LFS), the normal current density is negative and therefore it might be   eventually limited by the electron saturation current which is a factor $\sqrt{m_i/m_e}\approx 61$ larger than the ion saturation current. The limit to negative current in the LFS is not driven by the electron saturation current but by the limit to the ion saturation current on the HFS and electric charge conservation of the plasma, which leads to a similar negative current to $J_{sat}$ on the LFS despite the $\Phi = 0$ boundary condition there.

\medskip 
 
  Note that in the green shaded region of Figure \ref{fig:Jlim_proof} the halo current is somewhat larger than the ion saturation current in the simulations. Unfortunately, the current could not be limited near the contact point or limiter point (green region) for the following reasons. Around the contact point (defined by $\v{B}\cdot\v{n}=0$), large gradients of the electrostatic potential are formed along the boundary ($\grad_\textrm{tan}\Phi$). The latter is due to the fact that at exactly the contact point $\Phi=0$, and in its vicinity, large voltages are applied in order to limit the parallel current (since $\v{B}\cdot\v{n} \neq 0$). Since the parallel current density can locally increase due to these gradients via the local Ohm's law ($J_\parallel\propto -\v{B}\cdot\grad\Phi/\eta$), an unstable situation can occur around this point. In this case, an increase of $J_\parallel$ near the limiter point raises $\v{B}\cdot\grad\Phi$, which in turn increases $J_\parallel$ even further. To avoid this issue, the following boundary condition is imposed to control  the gradient in $\Phi$ along the boundary

\begin{equation}
 \Phi = - D_{lim}\pderiv{\Phi}{l}
\end{equation} 
where $D_{lim}$ is the distance in real space to the contact point and $l$ is the length along the boundary. The latter boundary condition is applied up to 6 cm away from the limiter point, which is the minimum stable distance that has been found by a trial and error analysis. The effect of this boundary condition on the electrostatic potential can be observed in the green region of figure \ref{fig:Jlim_proof}. With this setup, it has been found that for the latter phases of the VDE (e.g. Figure \ref{fig:Jlim_proof}), the maximum normal current density ($\v{J}\cdot\v{n}|_\textrm{max}$) can exceed the maximum ion saturation current ($\v{J}_{sat}\cdot\v{n}|_\textrm{max}$) in this region by only up to $\sim$30\%  of its value.

\begin{figure}
\centering
  \includegraphics[width=0.65\textwidth]{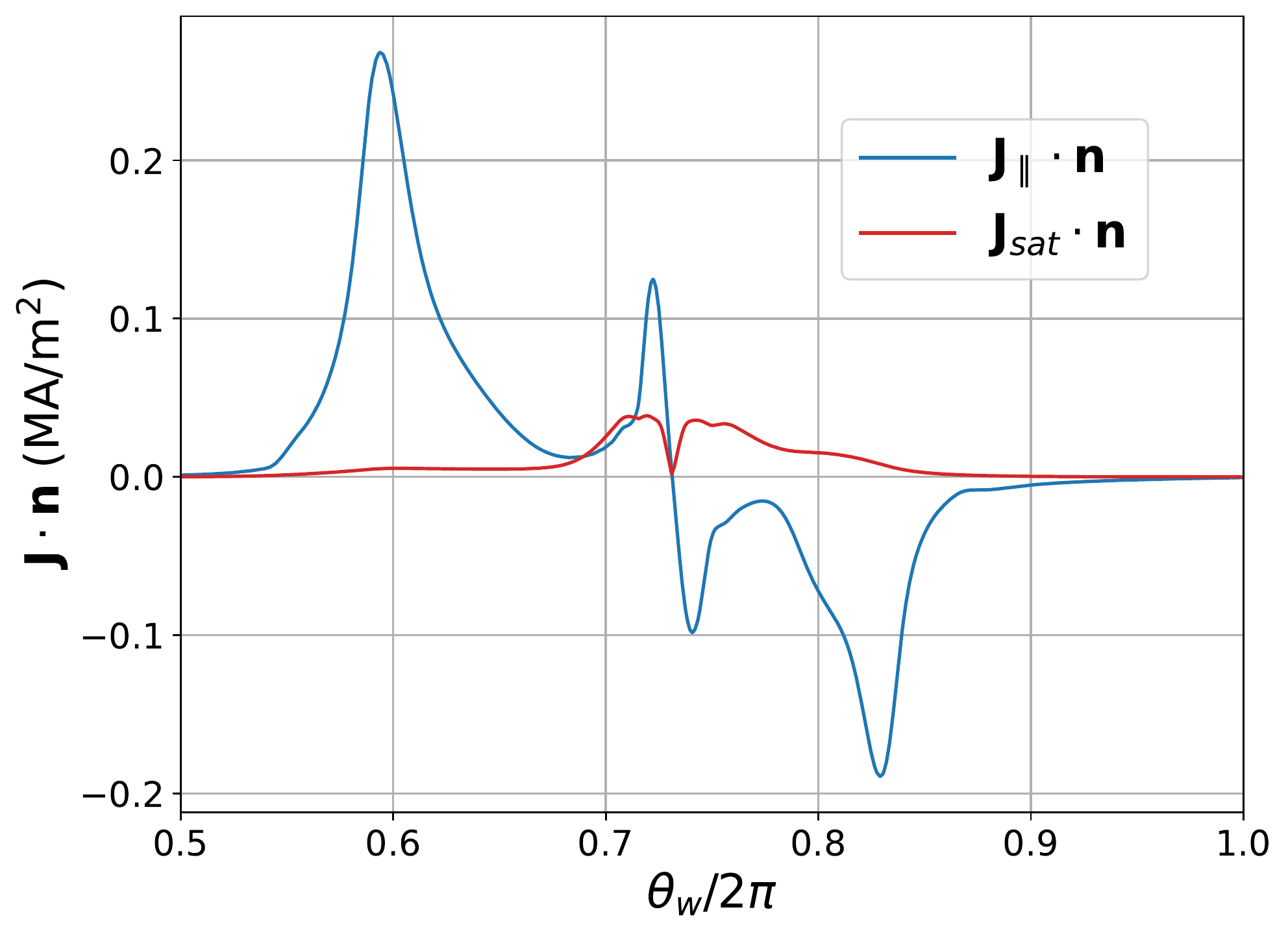}
\caption{ Normal current and ion saturation current density at the wall-plasma interface as function of the polar angle ($\theta_w$) for the simulation with $\Phi=0$ at the time when $Z_{axis}=-0.27$ m. The figure shows that the parallel current density becomes larger than the ion saturation current for the usual MHD boundary condition, violating the experimentally found physical limit.}
\label{fig:Jlim_violation}
\end{figure}

\begin{figure}
\centering
  \includegraphics[width=0.6\textwidth]{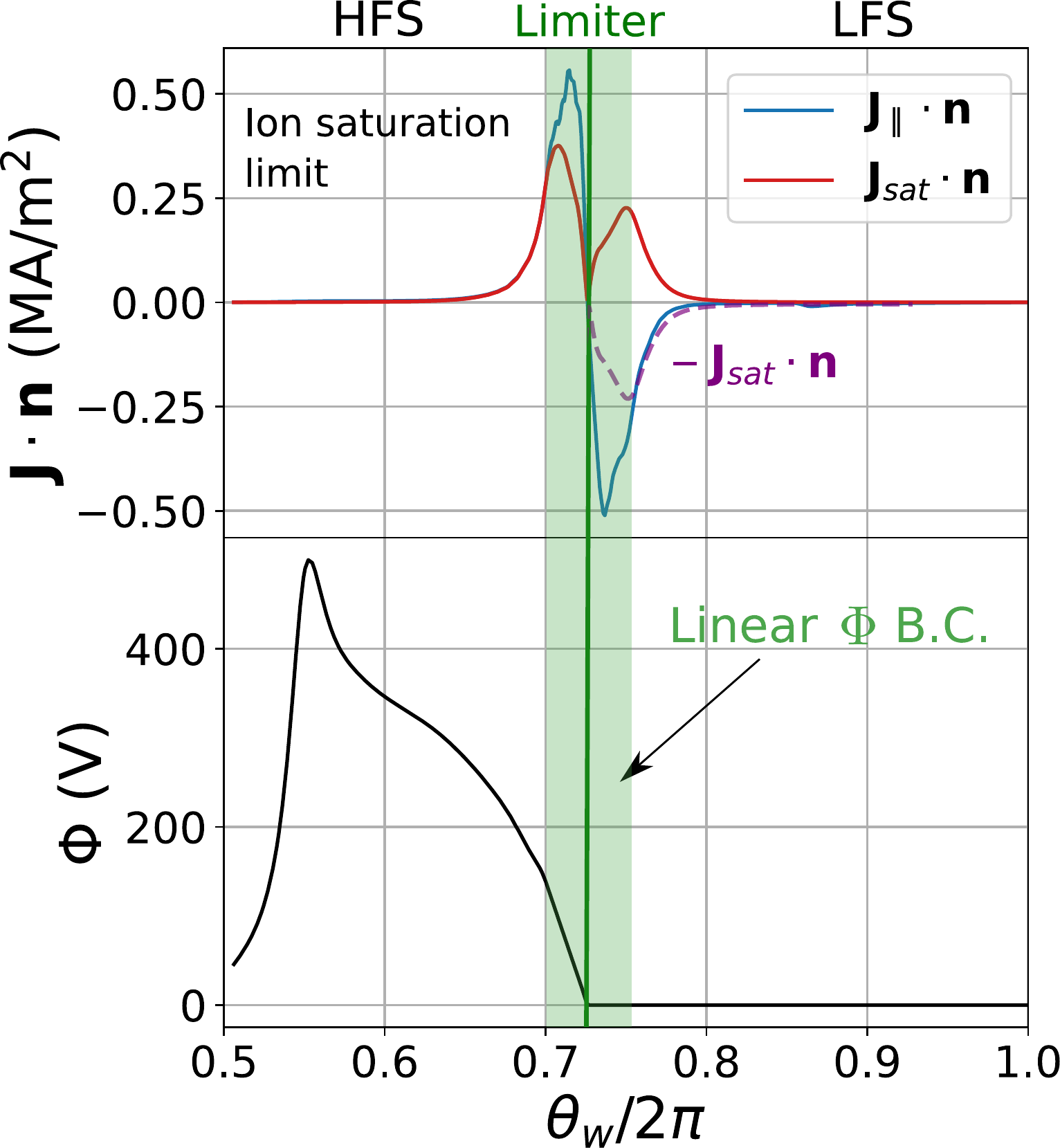}
  \caption{(Top) Normal current and ion saturation current density  at the wall-plasma interface as function of the polar angle ($\theta_w$) for the "Base" VDE simulation  at the time when $Z_{axis}=-0.27$ m. The limiter point is indicated by the green line and the green area is the region where linear boundary conditions are applied for $\Phi$. The purple dashed line is $-\v{J}_{sat}\cdot \v{n}$ at the Low Field Side  (see text for the physical limitations to negative current in the LFS). (Bottom) Electrostatic potential as a function of $\theta_w$.}
  \label{fig:Jlim_proof}
\end{figure}

\medskip

Finally we investigate the origin of the non-monotonic current profile for the case with $\Phi=0$. Assuming spatially constant $T_e$, $J_\parallel$ and $n$ and that the Ohmic heating and the parallel conduction are dominant, the evolution of the temperature in a SOL flux tube is
\begin{equation}
 \pderiv{T_e}{t} \approx \frac{\eta J_\parallel^2}{n} -  c_{sh} T_e^{3/2} \frac{A_{tube}}{V_{tube}} = \left(\eta_0 T_0^{-3/2}\frac{ E_\parallel^2}{n}-c_{sh} \frac{A_{tube}}{V_{tube}} \right) T_e^{3/2}
\end{equation} 
where $V_{tube}$ is the volume of the flux tube, $A_{tube}$ is the cross-sectional area of the flux tube at its end points, $c_{sh}\equiv 2\gamma_{sh}\sqrt{\gamma k_B^3/m_i}$ and we have used that $E_\parallel=\eta J_\parallel$. The latter equation shows that for a given $E_\parallel$  the flux tubes with smaller density and with smaller $A_{tube}/V_{tube}$ ratios are the most efficient ones for increasing $T_e$. The flux tubes in the upper part of the plasma domain are the most efficient driving temperature because of the low density and the low $A_{tube}/V_{tube}$ ratio. That is the reason why the current is preferentially  induced on those regions and a non-monotonic profile is observed in Figure \ref{fig:compare_u0} b).

\subsection{$q_{min}$ and $\Gamma_{min}$}
In order to test the influence of $q_{min}$ and $\Gamma_{min}$, two different simulations were run with $\vartheta_{min}=1^\circ$ and $\vartheta_{min}=2^\circ$. The influence of $\vartheta_{min}$ on the vertical position and current decay is not significant, but the $T_e$ near the contact point can be increased by $25\%$ when $\vartheta_{min}=1^\circ$ (see figure \ref{fig:qmin_test}).

\begin{figure}
\centering
  \includegraphics[width=0.5\textwidth]{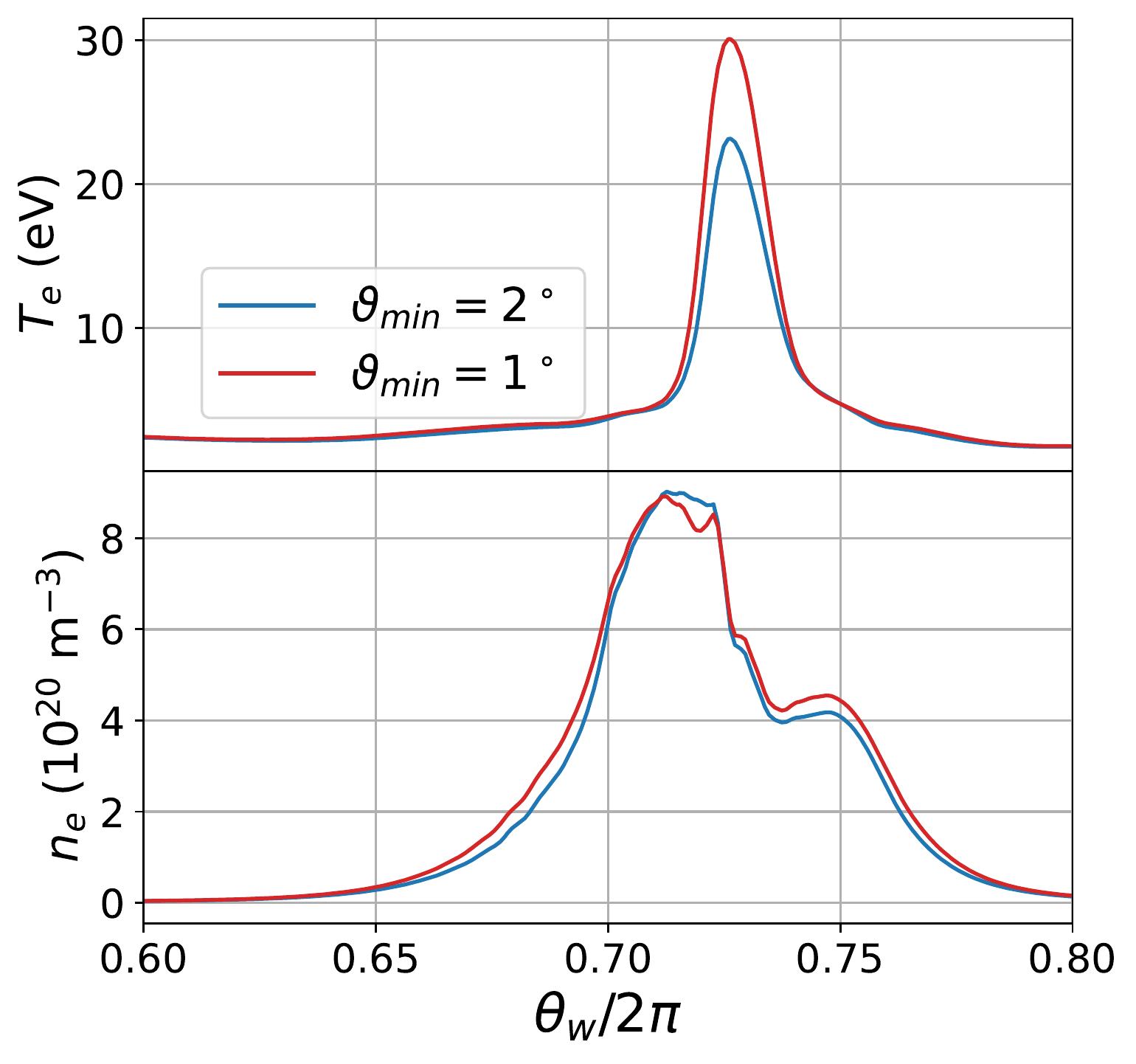}
\caption{$T_e$ and $n_e$ profiles along the wall for two different   $\vartheta_{min}$ values used in equations \eqref{eq:q_flux_BC} and \eqref{eq:part_flux_BC} at $I_p=0.7I_{p0}$.  }
\label{fig:qmin_test}
\end{figure}

\begin{figure}
\centering
  \includegraphics[width=0.9\textwidth]{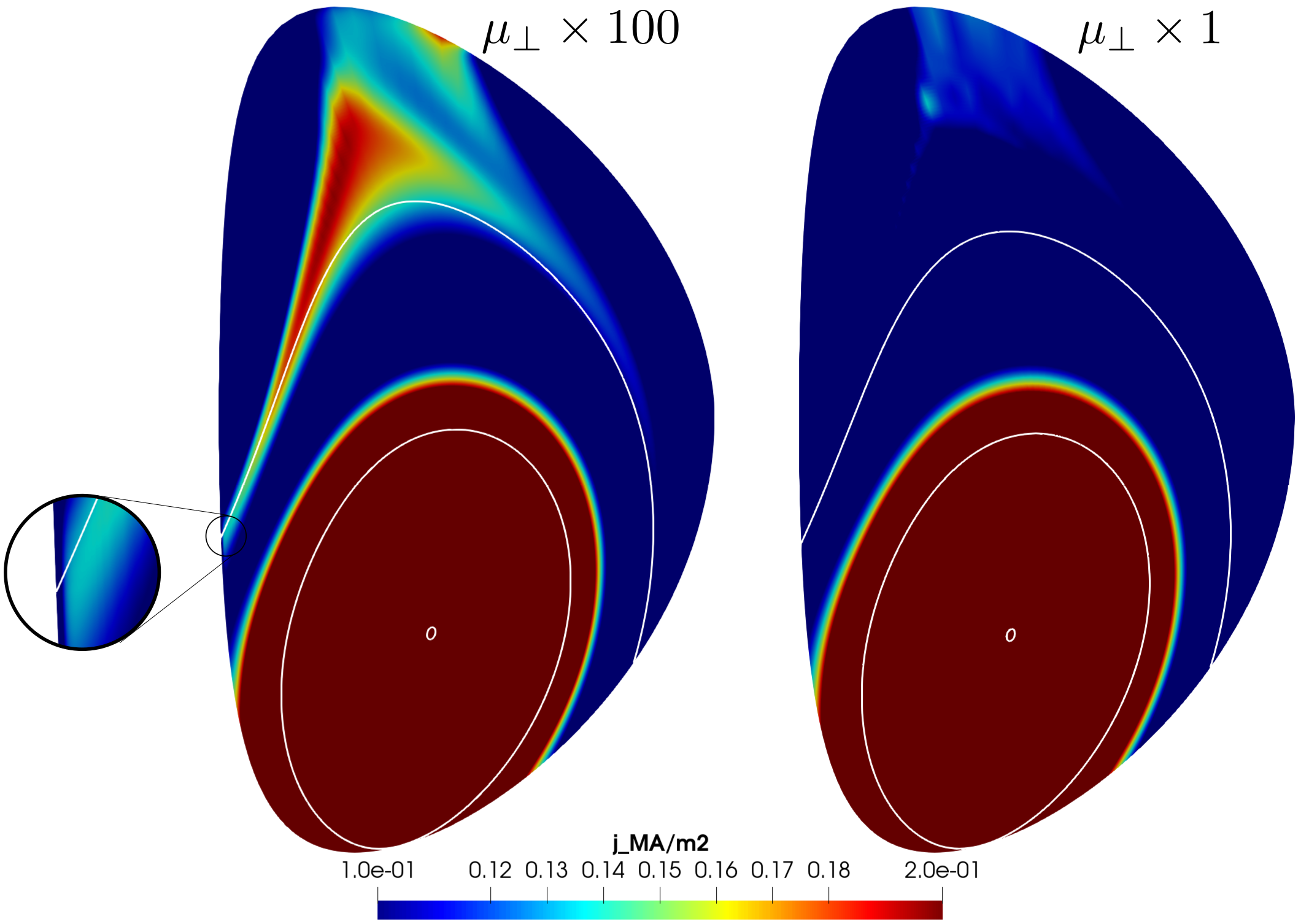}
\caption{Current density in the far SOL in (left) simulation with $\mu_\perp = 1.05 \times 10^{-5} \text{kg m/s}$ and (right) simulation with $\mu_\perp = 1.05 \times 10^{-7} \text{kg m/s}$. The viscosity coefficients are spatially constant.  The white contours indicate three different flux surfaces. When a large unphysical viscosity is present, current gradients along the field lines can be sustained.    }
\label{fig:visco_test}
\end{figure}

\subsection{Effect of viscosity on ion saturation current boundary condition}
The choice of the plasma perpendicular viscosity is found to play a very important role in the limitation of $J_\parallel$.  Near Alfvenic equilibrium and for cold plasmas, the current density is parallel to the magnetic field lines  ($\v{J}\times \v{B}=0$). In the absence of charge accumulation ($\div \v{J}=0$), the current density also satisfies 

\begin{equation}
\v{B}\cdot\grad (J_\parallel/B)\approx  \frac{1}{F_0}\v{B}\cdot\grad j = 0
\end{equation}
which implies that the quantity $j\equiv R J_\phi$ is approximately constant along the magnetic field-lines. When the boundary condition for $\Phi$ is applied, the current density at the boundary is locally limited by $J_{sat}$. This creates a gradient of current along field-lines leading to the excitation of Alfven waves which restore $\v{B}\cdot\grad j =0$. For time-scales longer than the Alfven time, gradients along the field-lines must vanish implying that  $J_\parallel \leq J_{sat}|_{wall}$. In the presence of unphysically large fluid viscosities, a balance between the viscous force and the current gradient force can be established ($\v{B}\cdot\grad j \approx \mu_\perp \Delta w$). Such a balance can prevent the excitation of Alfven waves and the homogenization of the current along the field-lines leading to $J_\parallel > J_{sat}|_{wall}$. A case with large viscosity is presented in figure \ref{fig:visco_test} (left) showing the strong effect of the viscosity in the SOL current profile and the existence of large current density gradients along field-lines. Therefore special care must be taken with the viscosity choice when applying this type of boundary conditions. To prevent unphysical results in our simulations, the value of $\mu_\perp$ has been chosen small enough to avoid gradients of current density and such that a smaller value does not influence the plasma dynamics.

\begin{figure}
\centering
  \includegraphics[width=0.47\textwidth]{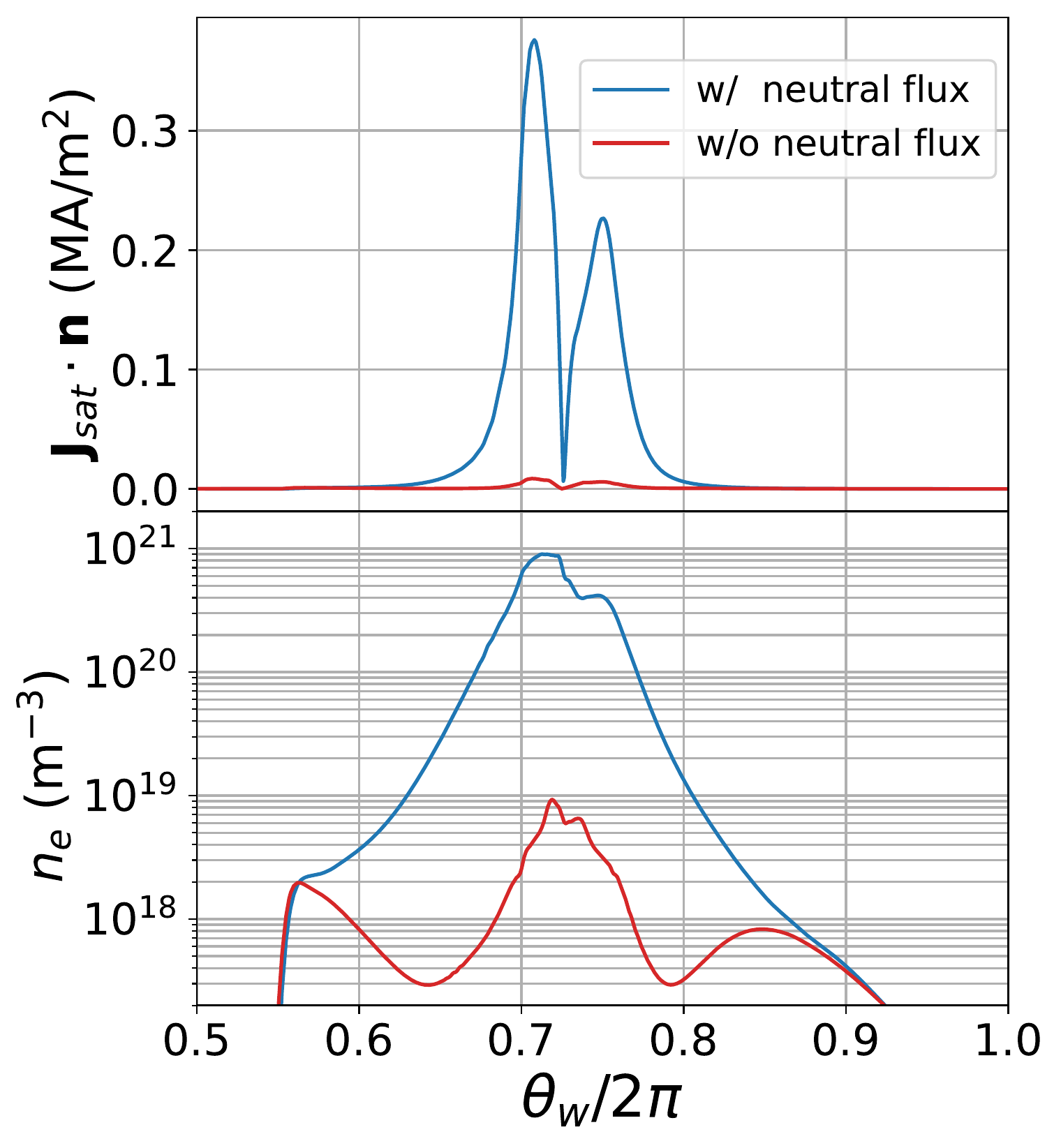}
\caption{ Normal ion saturation current density (top) and electron density (bottom) at the wall-plasma interface as function of the polar angle ($\theta_w$) for the time when $Z_{axis}=-0.27$ m. The blue line corresponds to the Base VDE simulation and the red line to a VDE case in which the neutral particle flux has been set to zero ($\v{\Gamma}_n\cdot\v{n} =0$). The figure shows the importance of including neutral particles to calculate the electron particle density and the ion saturation current at the plasma-wall interface. }
\label{fig:neutral_effect}
\end{figure}

\subsection{Effect of the neutral particle flux}
In this subsection we discuss the large effect of  neutral particles on the electron density and on the ion saturation current. The VDE simulation was restarted from the end of the steady state run and the neutral particle flux was to zero ($\v{\Gamma}_n\cdot\v{n} =0$). For that case, the electron density and the ion saturation current is compared to the Base simulation at a latter phase of the VDE in Figure \ref{fig:neutral_effect}. When the neutral particle flux is set to zero, the plasma ions leave the computational domain at the sound speed and do not come back as neutral particles. Therefore the boundary acts as a large ion sink that leads to plasma densities lower than the Base simulation by two orders of magnitude (see Figure \ref{fig:neutral_effect}). As a consequence, the ion saturation current density is strongly reduced and  halo currents cannot be induced during the VDE. The latter demonstrates that neutral particles must be included when using sheath boundary conditions in order to a obtain a self-consistent evolution of the plasma density and the ion saturation current at the plasma-wall interface.

\section{Comparison with divertor probe measurements}
\label{sec:comparison}

\begin{figure}
\centering
  \includegraphics[width=0.47\textwidth]{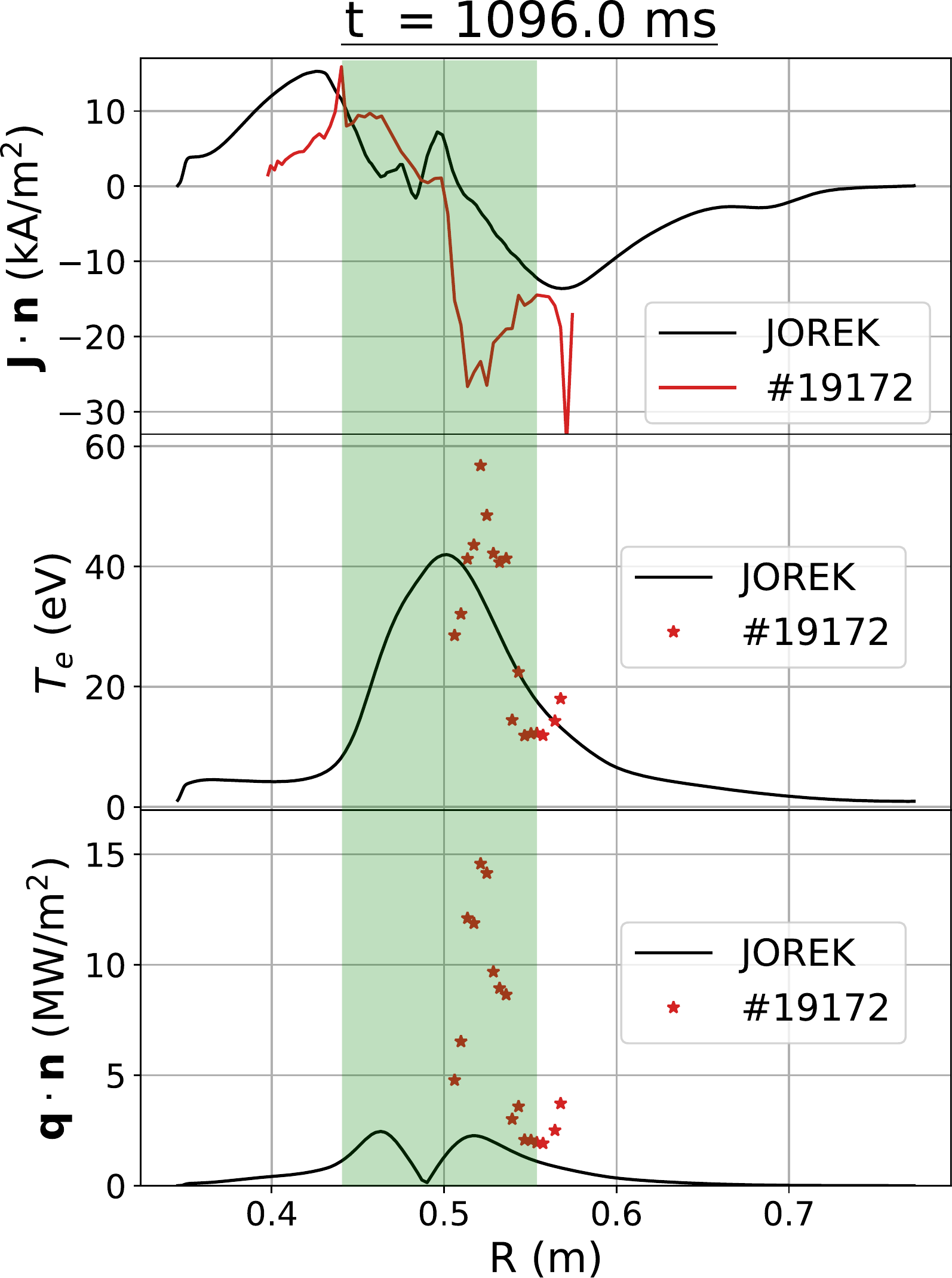}
  \includegraphics[width=0.48\textwidth]{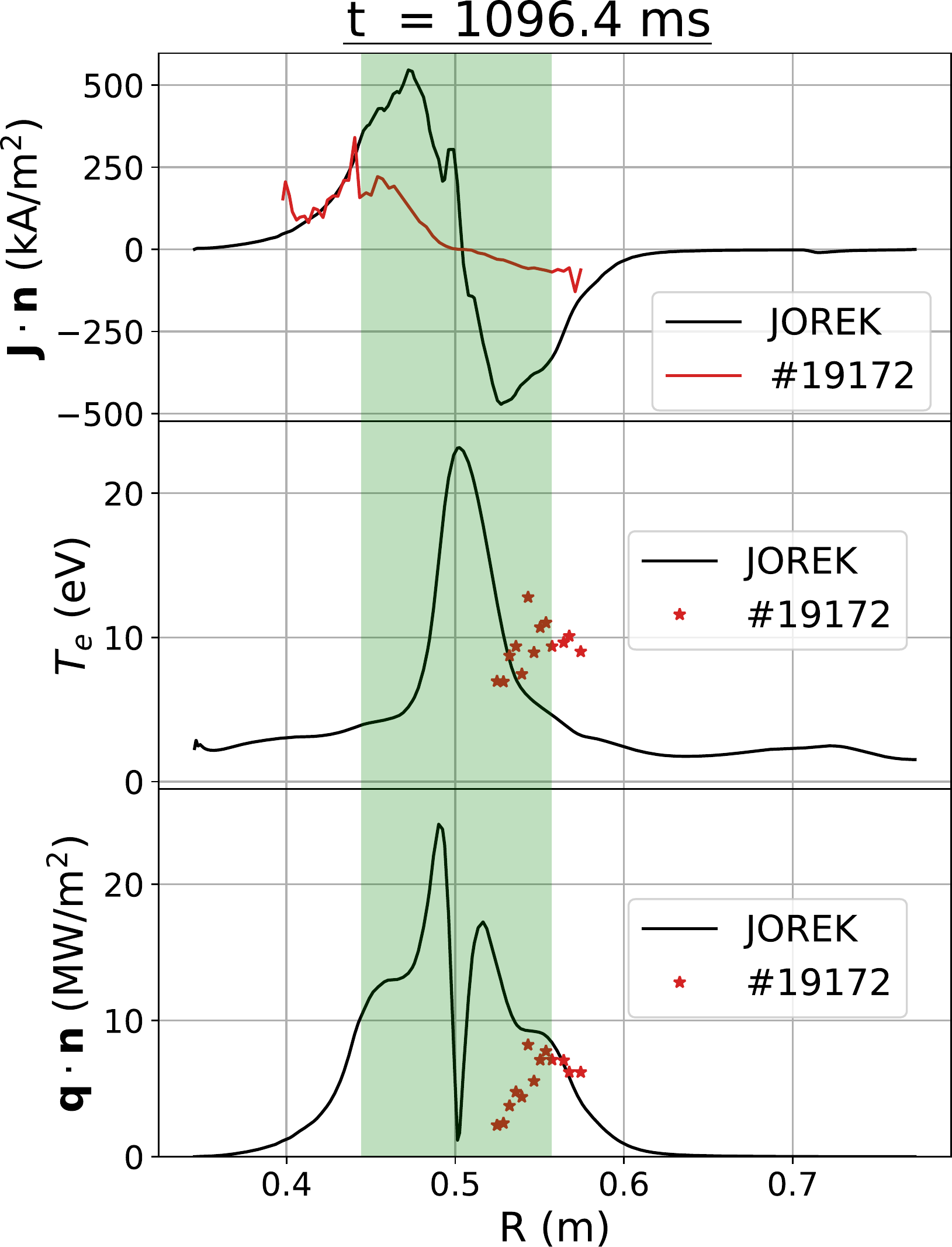}
\caption{ Normal current density, electron temperature and normal heat flux on the divertor at (left) $t=1096.0$ ms and at (right) $I_p=0.7I_{p0}=0.21 $ MA ($t=1096.40$ ms in the experiment and $t=1096.38$ ms in JOREK). The JOREK results for the Base VDE simulation are shown in black and the experimental measurements in red. The experimental heat flux has been estimated with $q_\parallel \approx 11 T_e |J_\parallel|$ and $\gamma_{sh}=11$. The shaded  region in green corresponds to the area where the $J_{sat}$ limitation of the current density is not effective and a linear boundary condition is applied for the electrostatic potential  as explained in section \ref{sec:influence_BC_Phi}. }
\label{fig:compare_exp2}
\end{figure}

For the COMPASS experimental campaign on VDEs and current flows, two arrays of rooftop-shaped Langmuir probes (LPs) and one array of Ball-pen probes (BPPs)  \cite{Adamek_2017} were used to measure the parallel current density and the electron temperature at the lower divertor. The location of the LPs array along the divertor is indicated in figure \ref{fig:eq_plot} (c). Since during COMPASS disruptions it is found that  $J_\parallel\approx J_{sat}$, the experimental heat-flux can be estimated by $q_\parallel = \gamma_{sh} T_e J_{sat} \approx \gamma_{sh} T_e |J_\parallel|$. Note however that this estimation is given for reference and that $\gamma_{sh}$ is assumed to be 11 while it could vary significantly during the disruption since strong halo currents circulate in the SOL. In figure \ref{fig:compare_exp2}, the JOREK values are compared to the experimental results along the boundary  at two different times. 

\medskip

 After a downwards motion of 10 cm and before the decay of the plasma current ($t=1096.0$ ms), the simulation shows current and temperature values in a similar range to those observed in the experiments (figure \ref{fig:compare_exp2} left). The location of the contact point in the simulation ($R_{lim}=0.490$ m) is in good agreement with the experimental location defined by $J_\parallel=0$ ($R_{lim}=0.499$ m). 
 The JOREK current profile is broader in the HFS than in the experiment, but due to the short width of the LP array it is not possible to conclude the same at the LFS. The large power flux in the experiment compared to JOREK is most likely due to the fact that the thermal quench is taking place at 1096.0 ms in the experiment (as indicated by $H_\alpha$ measurements) while such a quench is not included in this axisymmetric JOREK simulations.

\medskip

In figure \ref{fig:compare_exp2} (right) the same quantities are compared at the time where the total current reaches $70\%$ of its pre-disruptive value in the experiment and in the simulation. Excellent agreement is found on the location of the limiter position ($R_{lim}=0.50$ m). This particular time point is chosen because the current densities reach their maximum values $\sim 2$ MA/m$^2$. The experiments indicate that at this point, the temperatures decay to $\sim 10$ eV and that the heat-flux is significantly reduced. For the JOREK Base simulation, a thermal quench was not triggered artificially and cannot occur self-consistently due to the axisymmetric simulation setup, thus at this point, the plasma core still contains $28\%$ of its pre-disruptive thermal energy. Due to the large pressures  around the limiter point, the JOREK results show large heat fluxes compared to the experiments. However, where $J_\parallel$ is limited by $J_{sat}$ in the simulation, similar current densities and temperatures are found, although the JOREK simulations show that $T_e$ falls with a smaller decay length. The influence of several simulation parameters onto the results is shown in the next section.

\begin{figure}
\centering
  \includegraphics[width=0.47\textwidth]{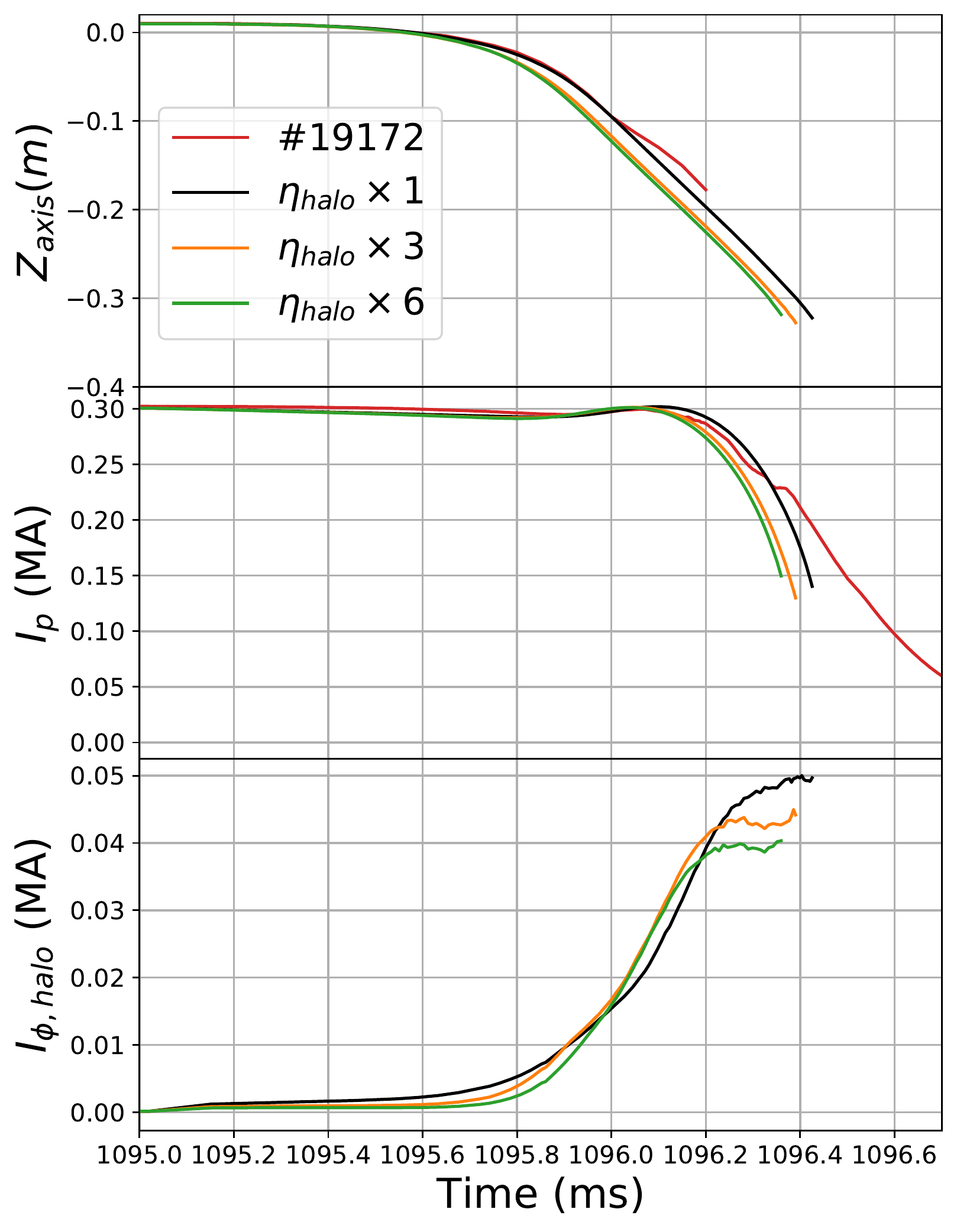}
  \includegraphics[width=0.49\textwidth]{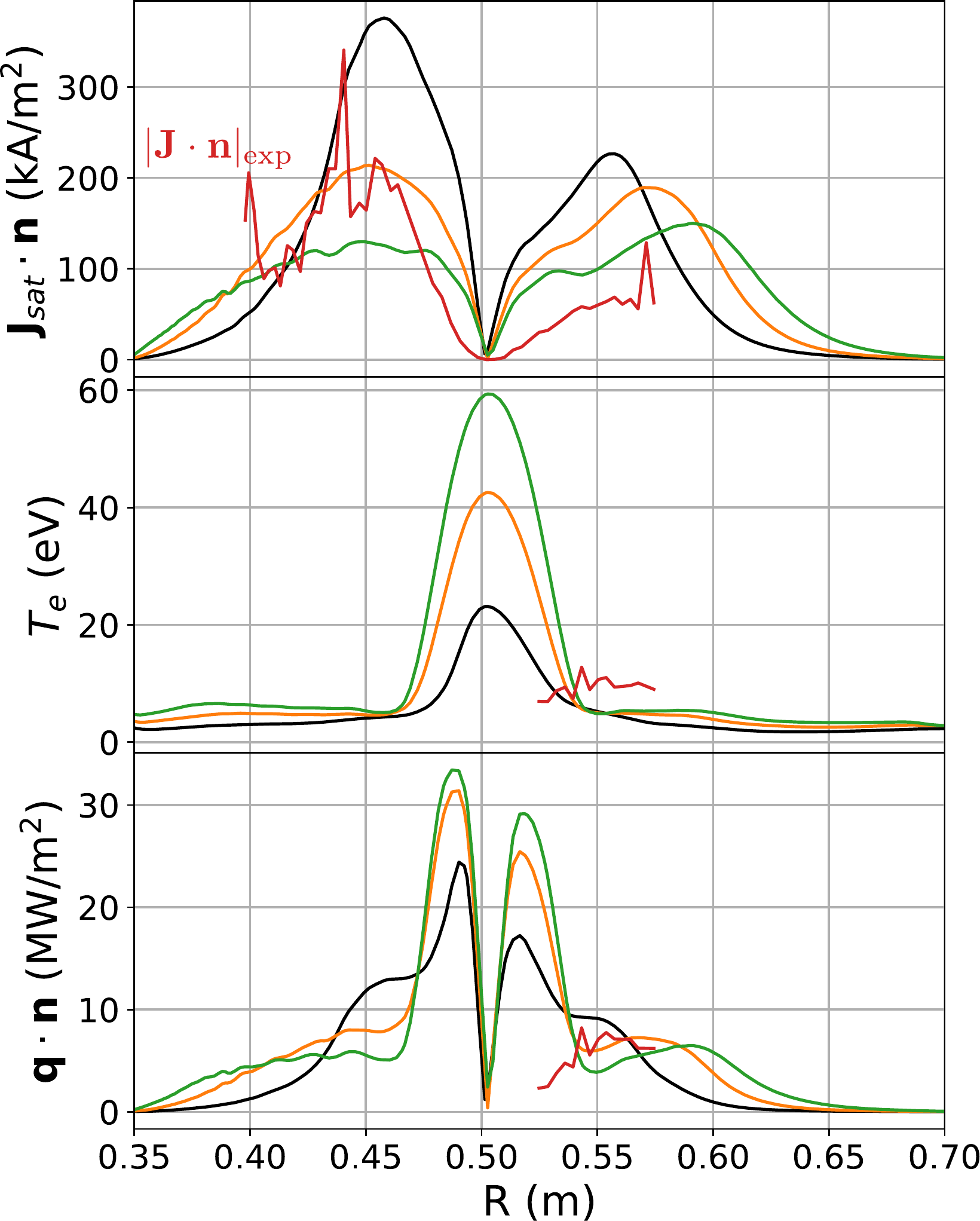}
\caption{ (Left) Vertical position of the magnetic axis, total toroidal plasma current and toroidal halo current in the plasma as function of time. (Right)  Normal ion saturation current density, electron temperature and normal heat flux on the divertor at $I_p=0.7I_{p0}=0.21 $ MA as function of the major radius. The lines with different colors correspond to simulations with different factors multiplying the plasma resistivity in the halo region. Note that in the upper right figure the absolute value of the experimental current density (red line) is plotted for reference. }
\label{fig:eta_halo_scan}
\end{figure}

\begin{figure}
\centering
  \includegraphics[width=0.47\textwidth]{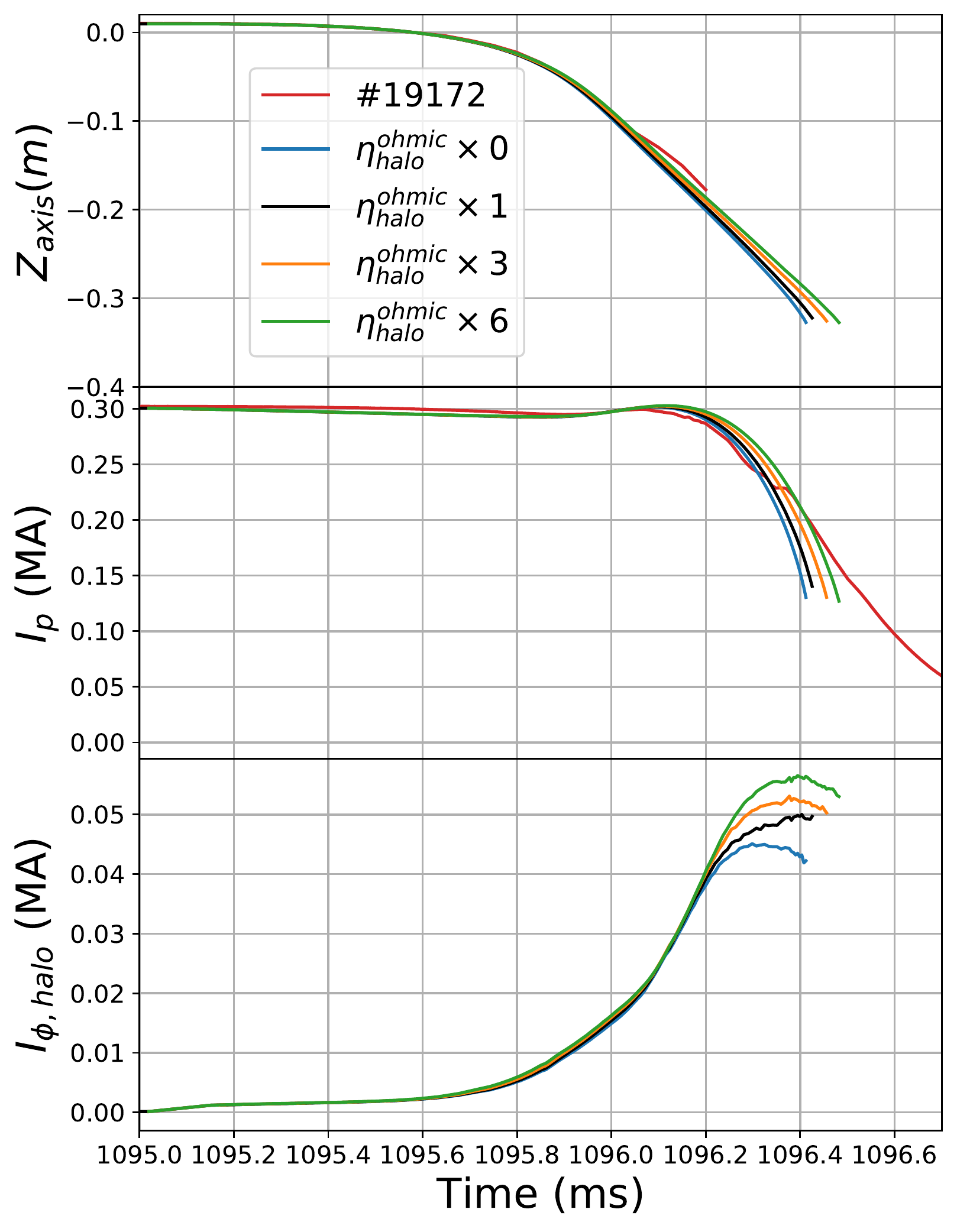}
  \includegraphics[width=0.49\textwidth]{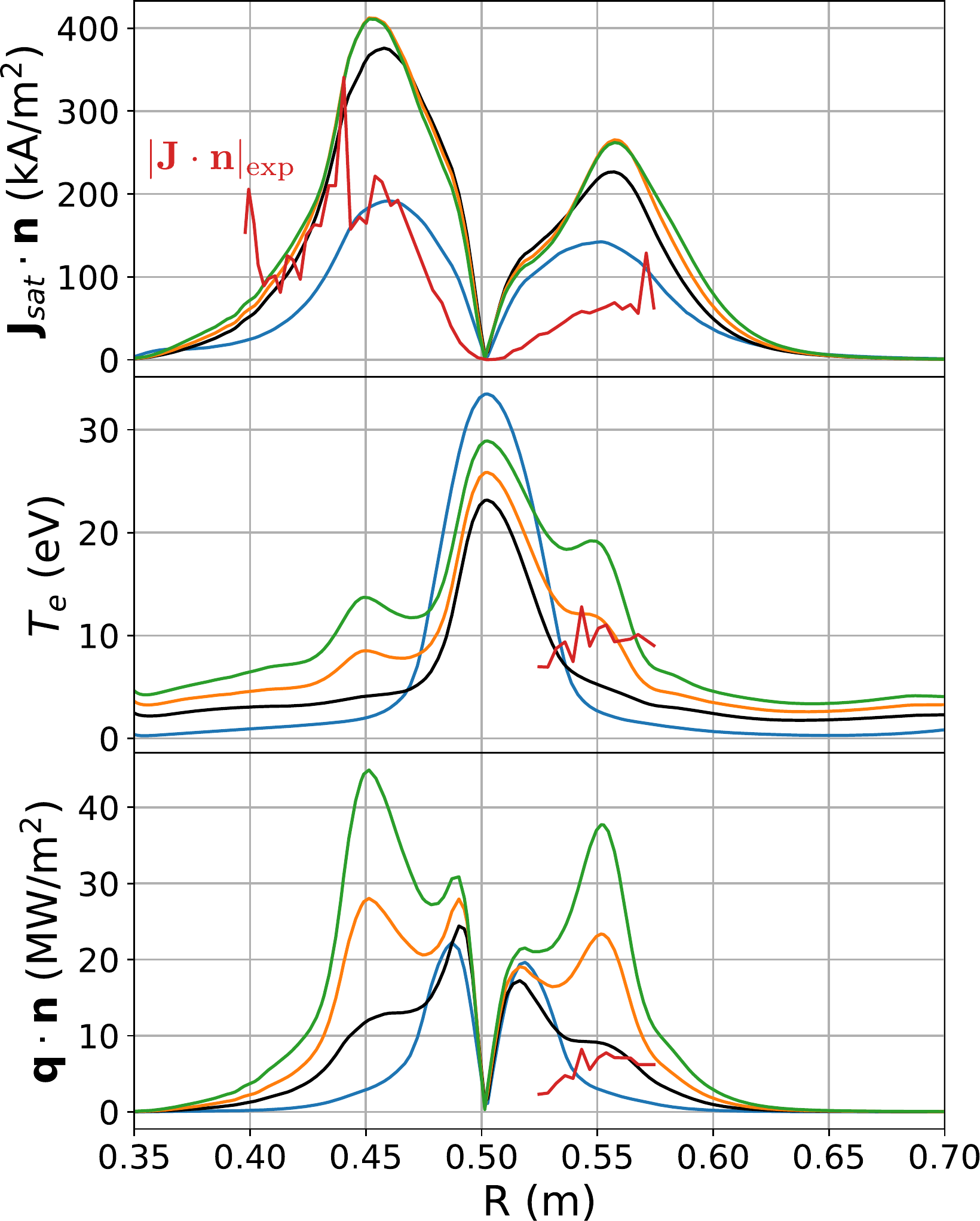}
\caption{ (Left) Vertical position of the magnetic axis, total toroidal plasma current and toroidal halo current in the plasma as function of time. (Right)  Normal ion saturation current density, electron temperature and normal heat flux on the divertor at $I_p=0.7I_{p0}=0.21 $ MA as function of the major radius. The lines with different colors correspond to simulations with different factors multiplying the Ohmic heating term in the halo region without modifying current diffusion. Note that in the upper right figure the absolute value of the experimental current density (red line) is plotted for reference. }
\label{fig:eta_ohmic_scan}
\end{figure}

\section{Sensitivity to different parameters}
\label{sec:sensitivity}
In this section we study how the different parameters shown in table \ref{tab:param_steady} influence the profiles at the divertor region. During all simulated cases, the parallel current was always found to be limited by the ion saturation current (except for $R\in [R_{lim}-0.06, R_{lim}+0.06]$). Therefore when analyzing the current density profile  we focus on how different parameters can affect the amplitude and shape of the $J_{sat}$ profile.

\subsection{Influence of resistivity and Ohmic heating}
The plasma resistivity dictates the diffusion of the current density and influences the energy balance through Ohmic heating. In figure \ref{fig:eta_halo_scan} we study the influence of the resistivity in the halo region by increasing it by a given factor only in that domain. Such an exercise can be viewed as a scan of the effective charge ($Z_{eff}$) in the halo region. In this case the resistivity is modified in both the induction equation ($\eta J_\parallel$ term) and in the pressure equation ($\eta J_\parallel^2$ term). The time traces in Figure \ref{fig:eta_halo_scan} (left) indicate that increasing $\eta_{halo}$ causes the vertical plasma motion and the current decay to take place earlier in time, although the decay rates do not  change significantly. Also the toroidal halo current has a weak dependence on $\eta_{halo}$. The profiles at the time when $I_p=0.7I_{p0}$ (Figure \ref{fig:eta_halo_scan} right) show that the increase in $\eta_{halo}$ flattens the $J_{sat}$ profile as well as the temperature and heat flux profiles. 

\begin{figure}
\centering
  \includegraphics[width=0.47\textwidth]{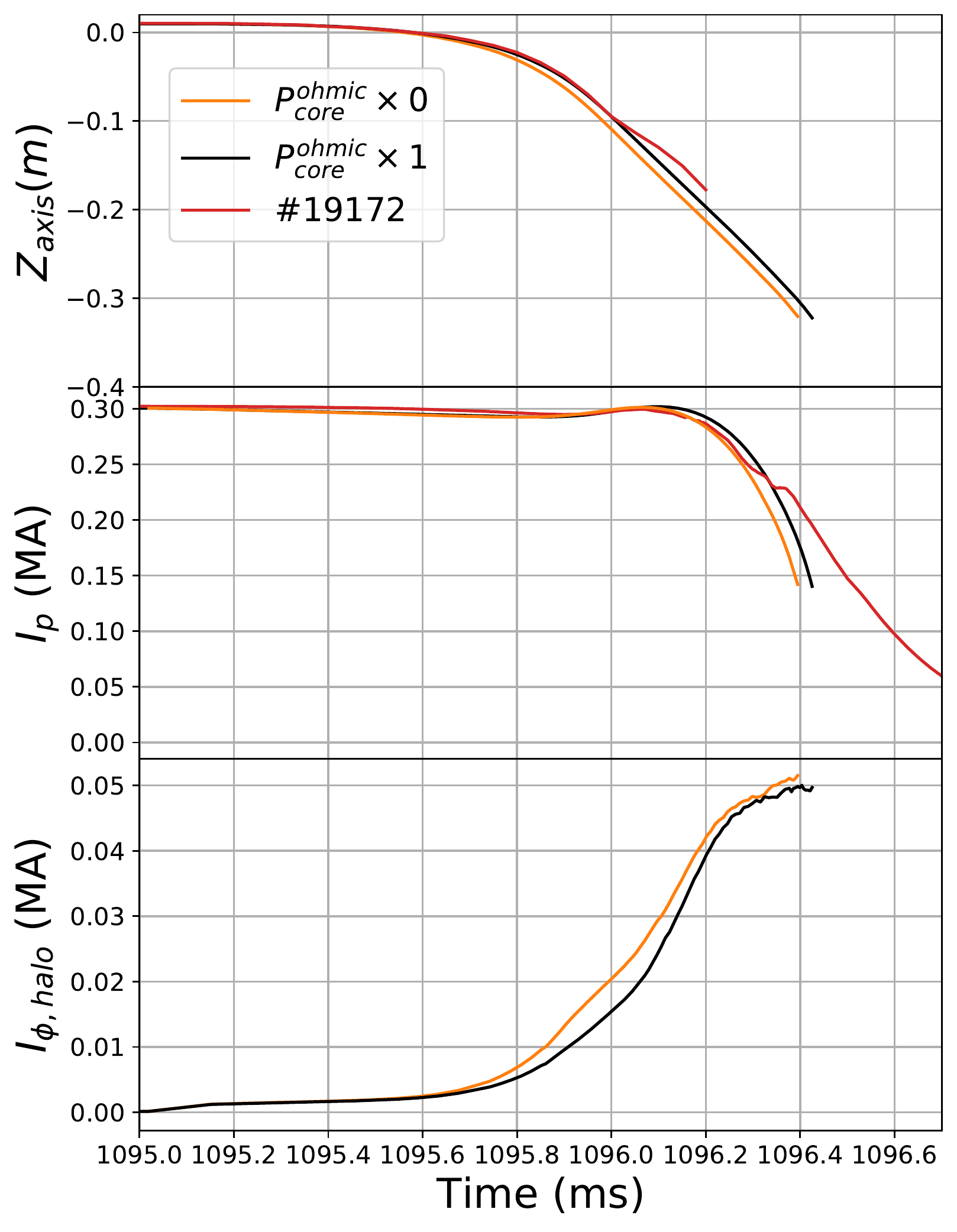}
  \includegraphics[width=0.49\textwidth]{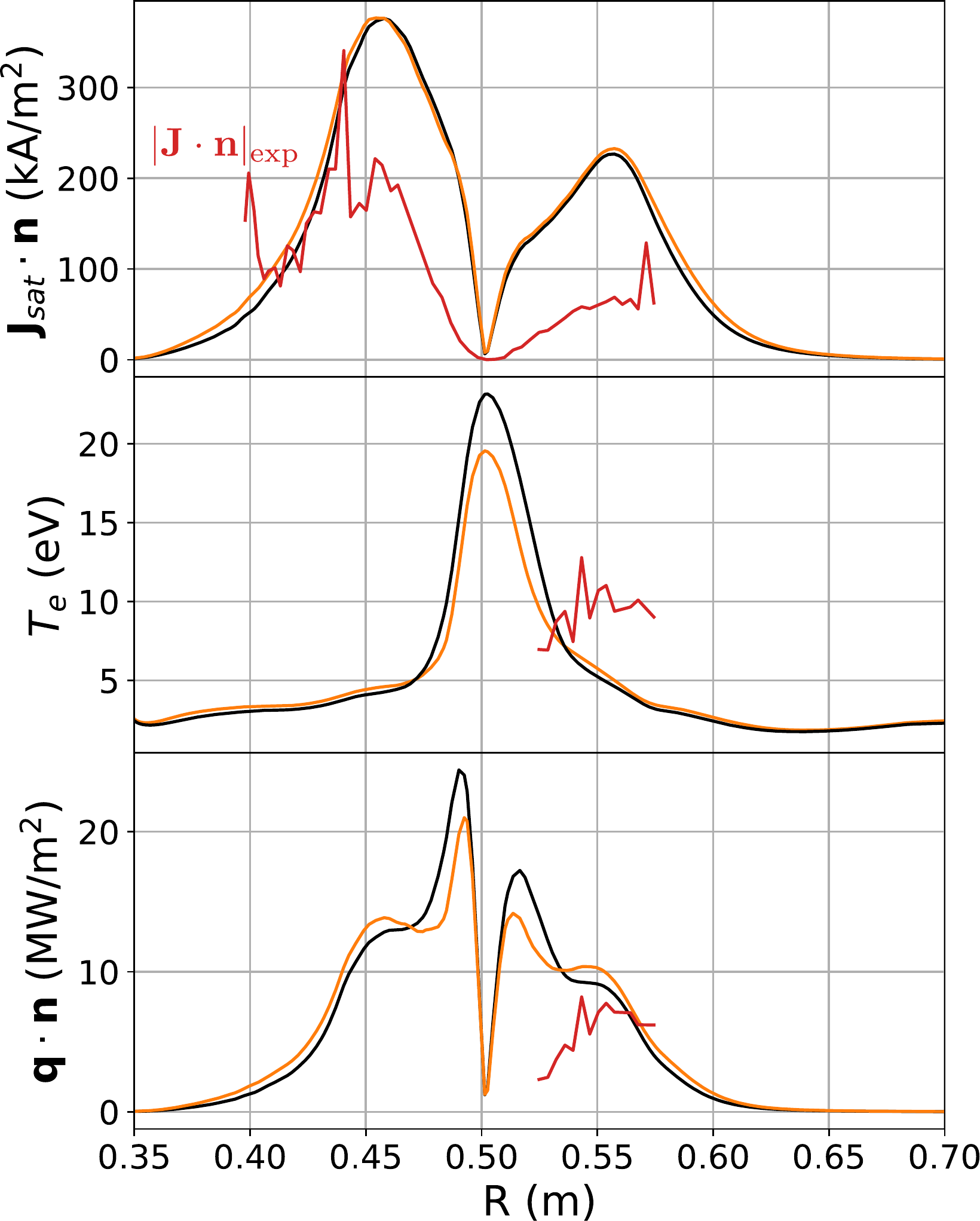}
\caption{(Left) Vertical position of the magnetic axis, total toroidal plasma current and toroidal halo current in the plasma as function of time. (Right)  Normal ion saturation current density, electron temperature and normal heat flux on the divertor at $I_p=0.7I_{p0}=0.21 $ MA as function of the major radius. The black line corresponds to the base simulation and the yellow line to a case with the core Ohmic heating switched off. Note that in the upper right figure the absolute value of the experimental current density (red line) is plotted for reference. }
\label{fig:core_ohmic_scan}
\end{figure}

In order to distinguish between the effect of the resistive diffusion and the effect Ohmic heating, a set of cases were run where only the Ohmic heating term was multiplied by different factors in the halo region. These cases are presented in Figure \ref{fig:eta_ohmic_scan}, where it can be seen that the Ohmic heating alone can effectively increase $T_e$ and $q$  in the halo region.  The increased Ohmic power acts as an additional heating source that is able to increase the plasma temperature and due to this also the pressure. For the factor range 1-6, the effect of the Ohmic power on the $J_{sat}$ profile is very weak, which means that the halo pressure is due mostly through a temperature increase and thus at an almost constant particle flux (note that $J_{sat}\sim p /\sqrt{T}$). However when neglecting completely the Ohmic heating in the halo region (blue line), the $J_{sat}$ magnitude is approximately reduced by a factor of 2. Note that in this case, the total toroidal halo current is not rigidly limited to $J_{sat}$ in the $\pm$ 6 cm near the tangency point due to numerical reasons, as described before, and this allows a significant amount of halo current to flow in this region.  To conclude, the effect of the Ohmic heating on the $J_{sat}$ profile for $Z_{eff}\geq 1$ is very weak, but if the term is not included, the profile's amplitude can be reduced significantly. Finally this exercise proves that the profile flattening observed in figure \ref{fig:eta_halo_scan} with increasing resistivity originates from current diffusion and not from Ohmic heating.

\medskip

The effect of the heating power crossing the LCFS on the profiles along the divertor is studied in figure \ref{fig:core_ohmic_scan}. We remind that in these simulations no auxiliary heating is employed, and all heating power comes from Ohmic heating. A case with the Ohmic heating turned off in the core (blue curve) shows that the core power has a small effect on the divertor profiles when compared to the standard case (black line) as it would be expected since the power flowing in the halo from the core is much smaller than the power directly deposited in the halo by Ohmic heating. For example for the Base simulation at $t=1096.38$ ms (70\% of $I_{p0}$), the power deposited by Ohmic heating in the core is 1.1 MW  and the power directly deposited in the halo region by Ohmic heating is 7.8 MW. 

\begin{figure}
\centering
  \includegraphics[width=0.47\textwidth]{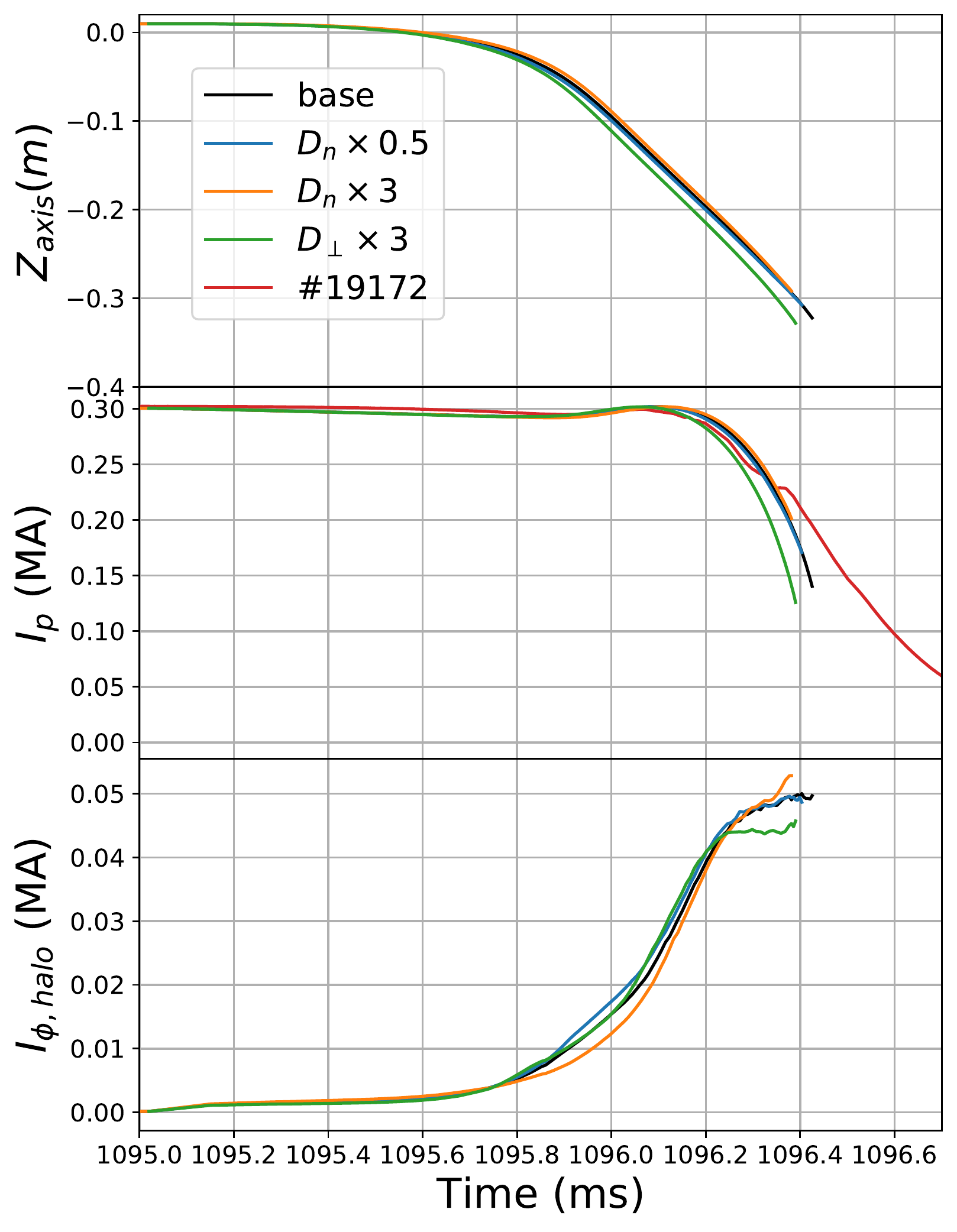}
  \includegraphics[width=0.49\textwidth]{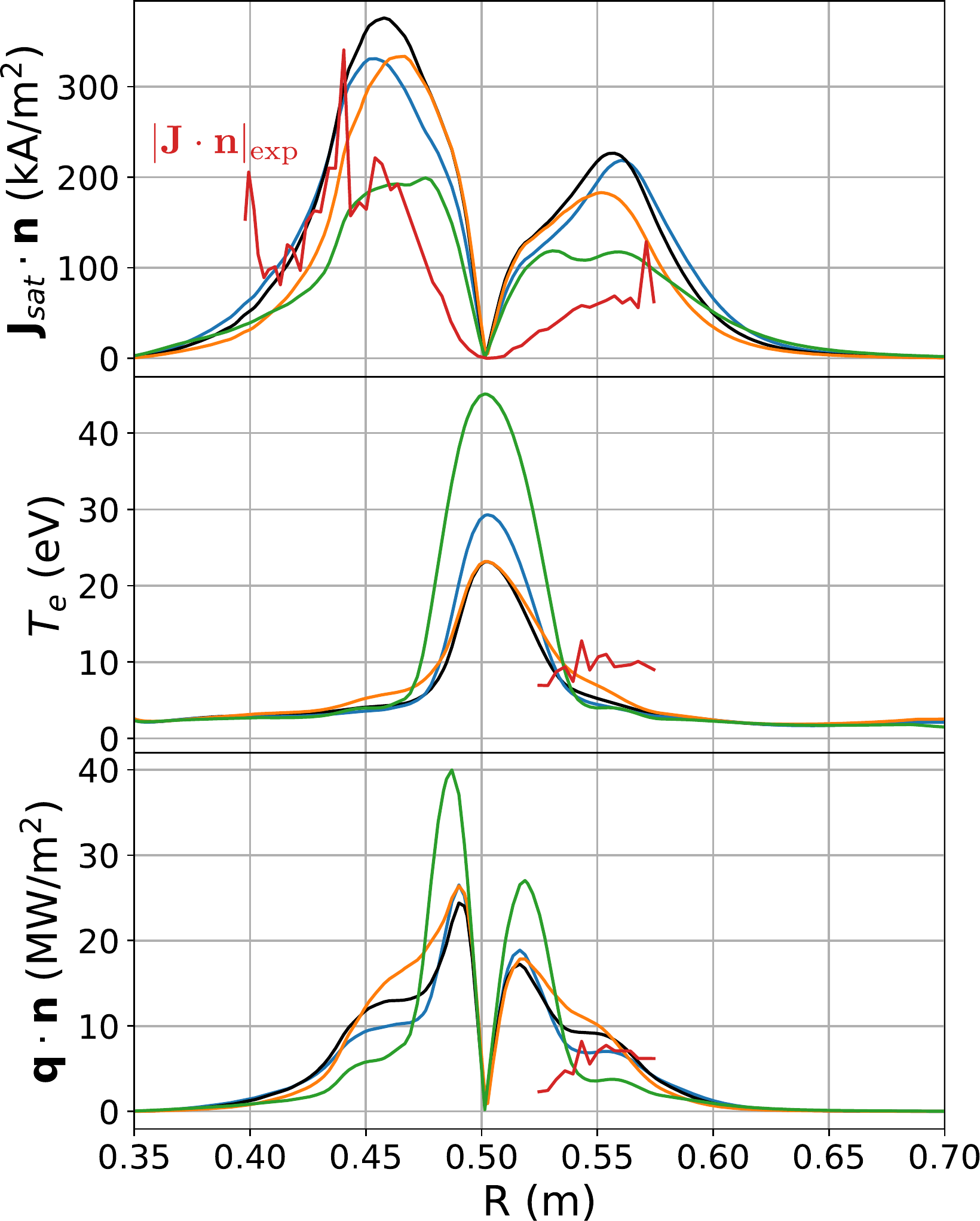}
\caption{ (Left) Vertical position of the magnetic axis, total toroidal plasma current and toroidal halo current in the plasma as function of time. (Right)  Normal ion saturation current density, electron temperature and normal heat flux on the divertor at $I_p=0.7I_{p0}=0.21 $ MA as function of the major radius. The legends correspond to a case with $D_n$ increased by a factor of 3 in all the plasma (green line), a case with $D_n$ decreased by a factor of 2 (yellow line), a case with the ion diffusion increased by a factor of 3 (blue line), the base simulation (black line)  and the experimental results (red line). Note that in the upper right figure the absolute value of the experimental current density (red line) is plotted for reference.}
\label{fig:particle_diff_scan}
\end{figure}

\begin{figure}
\centering
  \includegraphics[width=0.47\textwidth]{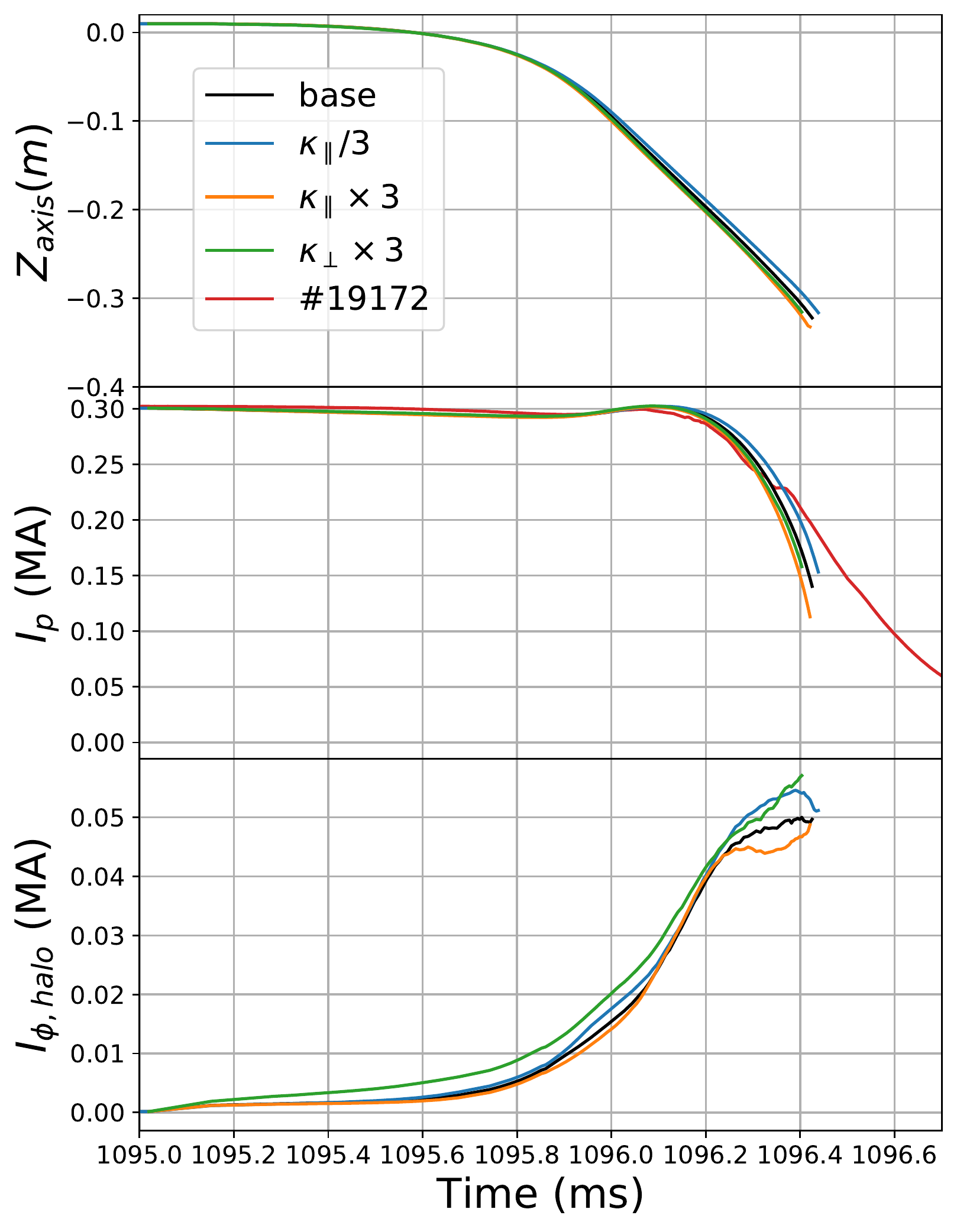}
  \includegraphics[width=0.49\textwidth]{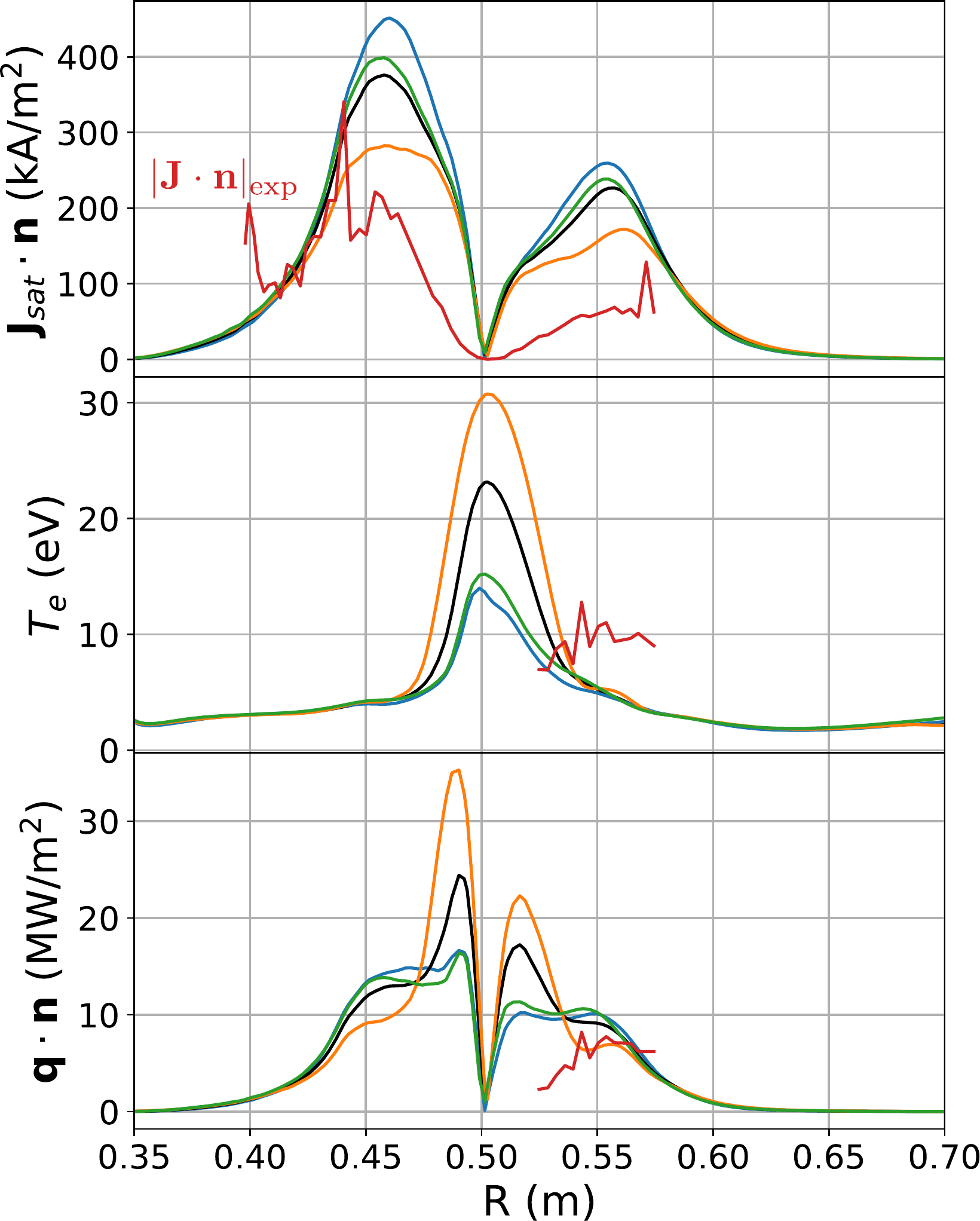}
\caption{ (Left) Vertical position of the magnetic axis, total toroidal plasma current and toroidal halo current in the plasma as function of time. (Right) Normal on saturation current density, electron temperature and parallel heat flux on the divertor at $I_p=0.7I_{p0}=0.21 $ MA as function of the major radius. The legends correspond to cases with $\kappa_\parallel$ decreased and increased by a factor of 3 (yellow and green curves), a case with $\kappa_\perp$ increased by a factor of 3 (blue line) the base simulation (black line)  and the experimental results (red line). Note that in the upper right figure the absolute value of the experimental current density (red line) is plotted for reference.}
\label{fig:heat_diff_scan}
\end{figure}

\subsection{Influence of particle and heat diffusion coefficients}
The particle diffusion coefficients can directly affect the distribution of the plasma density and therefore can modify the ion saturation current density. In Figure \ref{fig:particle_diff_scan} the standard simulation is repeated with different ion and neutral particle diffusion coefficients. When increasing the neutral particle diffusion coefficient by a factor of three (yellow line), the plasma density in the far SOL is reduced and the temperature increases, although the overall $J_{sat}$ profile is not strongly modified. A larger $D_n$ tends to decrease the  plasma density at the wall due to the increase of the neutral mean free path (i.e. the ion particle source locally decreases as the neutrals are ionized at a further distance from the wall). Decreasing this coefficient by a factor of two shows a small effect in the $J_{sat}$ profile, with a weak tendency to be  increased in the far SOL. The increase of the ion diffusion coefficient by a factor of three (green line) has a larger effect on the $J_{sat}$ profile. In this case the augmented $D$ can diffuse the plasma density at the wall and decrease $J_{sat}$ by about a factor of 2. Note that temperature does not increase accordingly and that finally the heat flux is reduced away from the contact point (e.g. at $R=0.45$ m). The reason is that a smaller $J_{sat}$ limiting the current flow away from the contact point, obliges the halo current to be re-induced near the LCFS and to deposit its associated Ohmic heating in that region instead.

\medskip

For steady state plasmas, the ratio between the perpendicular and the parallel heat conduction coefficients typically determines the width of the heat flux profile in the SOL. However for the considered VDE case, increasing the $\kappa_\perp /\kappa_\parallel$ ratio by a factor of three (either by modifying $\kappa_\perp$ or $\kappa_\parallel$) does not increase the width or the magnitude of the heat flux and the $J_{sat}$ profiles (see blue and green lines of Figure \ref{fig:heat_diff_scan}). This indicates that during VDEs the SOL width is determined by other processes such as the resistive current diffusion shown in the previous section. However the increase of the parallel heat conduction by a factor of three (yellow line of Figure \ref{fig:heat_diff_scan}) is able to reduce the $J_{sat}$ profile more significantly. In this case the divertor temperature near the limiter point raises as the upstream temperature is conducted faster along the field lines. As the LCFS region is hotter and therefore more electrically conducting, a significant fraction of the toroidal halo current is preferentially induced near the LCFS and deposits its associated Ohmic heating in that region, leading to the  decrease of the $J_{sat}$ magnitude at a further distance in the SOL (e.g. at $R=0.45$ m).

\section{Conclusions}
\label{sec:conclusions}
In this paper a fully self-consistent model for halo currents has been presented and explored numerically with the JOREK code. The reduced MHD model includes an equation for neutral particles and a set of advanced boundary conditions. A boundary condition for the electrostatic potential ($\Phi$) has been implemented in order to limit the halo current density ($J$) to the ion saturation current ($J_{sat}\sim n \sqrt{T}$) as expected from basic sheath plasma physics and observed in recent COMPASS experiments \cite{adamek2020}. The boundary condition for $\Phi$ requires the implementation of sheath boundary conditions for the parallel velocity, the ion density, the neutral density and the temperature in order to obtain a self-consistent  evolution of $J_{sat}$ at the plasma-wall interface. The choice of realistic plasma parameters (e.g. Spitzer $\eta (T)$ and Spitzer-Härm $\kappa _\parallel(T)$) together with this model extends the status of halo current simulations beyond the present state of the art. 

\medskip

The limitation of the halo current density to $J_{sat}$ is found to play a key role. The presented axisymmetric simulations of COMPASS VDE experiments show that with typical MHD boundary conditions ($\Phi=0$), the current density becomes much larger than the ion saturation current leading to non-physical results. When the boundary condition for the electrostatic potential is applied during the VDE, the halo current density is always limited by the ion saturation current $J\approx J_{sat}$ as observed in experiments, implying that the halo current density is mainly given by the particle density. In this case both the halo width and total halo current are reduced in comparison to the case in which the current density is not limited. The implementation of such boundary condition requires a special treatment  at the plasma-wall tangency point for numerical stability. In addition, the MHD simulations must be performed with a sufficiently low plasma viscosity in order to minimize current density gradients forming along the open field lines. The implementation of neutral particles is also found to be necessary when using sheath boundary conditions, otherwise the plasma-wall interface acts as a large ion sink that leads to very low particle densities and halo current densities. 

\medskip

Axisymmetric simulations of the COMPASS VDE discharge \#19172 were compared with experimental measurements. Time traces of the vertical position and the plasma current show a good agreement with experiments at the beginning of the current quench ($t=1096.0$ ms). A matching during the whole disruption is not expected as the axisymmetric modelling does not include yet important processes such as the thermal quench, impurity radiation and 3D MHD activity that can flatten the current density profile and affect the VDE dynamics. Our future  work will be focused on 3D MHD simulations with the presented model to include these processes in a self-consistent form. Comparisons with the divertor probe measurements show that the position of the limiter point during the VDE is well captured by the simulations. The measured divertor profiles  of the current density, the temperature and also the estimated heat flux were compared with the simulation results. The measured values are comparable in magnitude with the simulated values although significant discrepancies  can be observed at certain radial locations (differences can be larger than a factor of two).

\medskip

As the ion saturation current limit is found to determine the halo current density profile,  the effect on the parameter choice on $J_{sat}$ has been studied. In this respect the resistivity in the halo region flattens the $J_{sat}$ profile through resistive diffusion, leading to an increased halo width. The dependencies of $J_{sat}$ on the neutral particle diffusion and on the heat conduction coefficients are weaker than for the resistivity. However $J_{sat}$ is also sensitive to the ion diffusion coefficient, which can  reduce its magnitude considerably by reducing the plasma density at the boundary. In all the performed parametric studies the total halo current is only weakly affected by the specific parameter choice whether it affects or not $J_{sat}$. This implies that the current flow limit by $J_{sat}$ for the halo current density has a stronger impact on the halo width than on the total halo current with the latter being determined by VDE dynamics.  

\medskip

Assuming the same $J_{sat}$  limitation for halo currents during ITER VDEs, the ITER halo width will scale like $I_{p0}/(q_{95} \, J_{sat})$ since present experiments indicate that the total halo current scales with $I_{p0}/q_{95}$ \cite{Knight_2000,granetz1996disruptions}. Future simulations will be conducted with the presented model in order to determine whether the latter scaling is to be expected and how $J_{sat}$ evolves for ITER disruptions.

\section*{Acknowledgements}

This work was supported by the ITER Monaco Fellowship Program.  ITER is the Nuclear Facility INB no. 174. This paper explores physics processes during the plasma operation of the tokamak when disruptions take place; nevertheless the nuclear operator is not constrained by the results presented here. The views and opinions expressed herein do not necessarily reflect those of the ITER Organization. The simulations presented here have been performed using the Marconi-Fusion supercomputer. This work has been carried out within the framework of the EUROfusion Consortium and has received funding from the Euratom research and training program 2014-2018 and 2019-2020 under grant agreement No 633053. The views and opinions expressed herein do not necessarily reflect those of the European
Commission. The work was co-funded by MEYS projects number 8D15001 and LM2018117. This work has been carried out within the framework of the project COMPASS-U: Tokamak for cutting-edge fusion research (No. CZ.$02.1.01/0.0/0.0/16\_019/0000768$) and co-funded from European structural and investment funds.

\bibliography{biblio.bib}{}
\bibliographystyle{abbrvurl}

\end{document}